\documentclass[twocolumn,aps,prb,10pt,floatfix,superscriptaddress,longbibliography]{revtex4-2}
\usepackage{graphicx}
\usepackage{dcolumn}
\usepackage{bm}
\usepackage{amsmath}
\usepackage{amssymb}
\usepackage{siunitx}
\usepackage{color}
\usepackage{hyperref}
\renewcommand{\vec}[1]{\mbox{\boldmath$\mathrm{#1}$}}

\newcommand{\dd}{\mathrm{d}}

\begin{document}

\preprint{APS/123-QED}

\title{Interlayer and interfacial Dzyaloshinskii-Moriya interaction in magnetic trilayers: first-principles calculations}

\author{Tim Matthies}
\affiliation{Department of Physics, University of Hamburg, 20355 Hamburg, Germany}
\author{Levente Rózsa}
\affiliation{Department of Theoretical Solid State Physics, Institute for Solid State Physics and Optics, HUN-REN Wigner Research Centre for Physics, H-1525 Budapest, Hungary}
\affiliation{Department of Theoretical Physics, Institute of Physics, Budapest University of Technology and Economics, M\H{u}egyetem rkp. 3, H-1111 Budapest, Hungary}
\author{László Szunyogh}
\affiliation{Department of Theoretical Physics, Institute of Physics, Budapest University of Technology and Economics, M\H{u}egyetem rkp. 3, H-1111 Budapest, Hungary}
\affiliation{HUN-REN-BME Condensed Matter Research Group, Budapest University of Technology and Economics, M\H{u}egyetem rkp. 3, H-1111 Budapest, Hungary}
\author{Roland Wiesendanger}
\affiliation{Department of Physics, University of Hamburg, 20355 Hamburg, Germany}
\author{Elena Y. Vedmedenko}
\affiliation{Department of Physics, University of Hamburg, 20355 Hamburg, Germany}

\date{\today}

\begin{abstract}
We determine the Dzyaloshinskii-Moriya interaction within and between two magnetic cobalt layers separated by a non-magnetic spacer through \textit{ab initio} calculations.
We investigate different materials for the non-magnetic layer, focusing on the experimentally realized Co/Ag/Co system. We laterally shift the atoms in the non-magnetic layer to achieve the symmetry breaking required for the interlayer Dzyaloshinskii-Moriya interaction.
We compare the resulting interactions with the Lévy-Fert model and observe a good overall agreement between the model and the \textit{ab initio} calculations for the dependence on the atomic positions. Additionally, we derive a formula for the strength of the interlayer isotropic exchange interaction depending on the position of the atoms in the non-magnetic layer and compare it to the first-principles results. 
We investigate the limitations of the Lévy-Fert model by turning off the spin-orbit coupling separately on the non-magnetic and magnetic atoms and by studying the effect of band filling. Our work advances the understanding of the microscopic mechanisms of the interlayer Dzyaloshinskii-Moriya interaction and gives insight into possible new material combinations with strong interlayer Dzyaloshinskii-Moriya interaction.
\end{abstract}

\maketitle

\section{\label{sec:int}Introduction}
The Dzyaloshinskii-Moriya interaction~\cite{dzyaloshinsky1958thermodynamic,moriya1960anisotropic} (DMI) gives rise to non-collinear magnetic structures, like weak ferromagnets, chiral domain walls~\cite{vedmed2007chiraldomain} or magnetic skyrmions~\cite{romming2013writing}, and also influences their excitations. Consequently, it plays a crucial role in spintronics, spin-orbitronics, and magnonics~\cite{zhang2015magnetic,fert2013skyrmions,Garst2017}. 
In magnetic multilayers, there are two important contributions to this interaction: the interfacial DMI (IF-DMI) controlling the rotational sense of the spins within the layers~\cite{RoadMap} and the interlayer DMI (IL-DMI) controlling the rotational sense between the magnetic layers~\cite{vedmedenko2019interlayer,han2019long,arregi2023large,sandoval2024preservation}. Typically, the IL-DMI is much weaker than the IF-DMI~\cite{vedmedenko2019interlayer,han2019long,Avci2021,Liang2023}.  However, recently two systems with IL-DMI strength comparable to that of IF-DMI have been reported~\cite{arregi2023large,Yun2023}. 
The IF-DMI is primarily enhanced at interfaces of magnetic layers with non-magnetic materials possessing strong spin-orbit coupling (SOC), while it rapidly vanishes away from the interface. In contrast, the IL-DMI oscillates in strength as the thickness of the spacer layer is varied~\cite{Liang2023,arregi2023large}.
The IL-DMI modulates the spin structure when moving across the layers, giving rise to phenomena which cannot be explained by IF-DMI like chiral coupling between two ferromagnetic layers or unidirectional switching of the magnetization orientation with a preferred sense of rotation.
Recently, first successful applications of IL-DMI for field-free spin-orbit-torque magnetization switching have been reported by a number of groups~\cite{He2022,Liang2023,Lin2024,Wang2023,Xie2023,Li2023}, opening new perspectives in the area of three-dimensional spintronics.

The physical mechanisms behind the emergence of DMI include symmetry breaking and the presence of spin-orbit coupling. Moriya provided a complete set of symmetry rules together with a formula for the strength of the interaction in bulk magnets based on perturbation theory~\cite{moriya1960anisotropic}. Another explicit formula was proposed by Fert and Lévy in metallic spin glasses, where the DMI between two magnetic atoms is mediated by a non-magnetic impurity with strong SOC in a non-magnetic material~\cite{fert1980role,levy1981anisotropy}. Unfortunately, such perturbative formulae are unable to quantitatively predict the strength of the DMI as a function of material combination and layer thicknesses. Over the last decades, experimental efforts combined with first-principles calculations identified interfaces such as Co/Pt or Fe/Ir where a strong IF-DMI can be reliably observed. However, the value of the IF-DMI for the same material combination varies between different experimental and theoretical techniques~\cite{Kuepferling2023,Simon2018}. For the IL-DMI, the available data are more limited and many open questions remain.
Particularly surprising is the appearance of strong IL-DMI in magnetic multilayers with relatively light non-magnetic spacers like Ag, producing a rather weak IF-DMI~\cite{arregi2023large}. Another interesting aspect is the reported correlations between the thickness-dependent oscillations in the IL-DMI and those in the interlayer isotropic Ruderman–Kittel–Kasuya–Yosida (RKKY) interactions~\cite{Ruderman1954,Kasuya1956,Yosida1957,Liang2023,arregi2023large}. 
Existing theoretical investigations report wide-ranging values for the magnitude of the IL-DMI: for two Co layers separated by 1 to 3 non-magnetic layers, Ref.~\cite{han2019long} gives a maximal value of $\SI{2}{meV}$, Ref.~\cite{demiroglu2024oscillatory} reports $\SI{0.04}{meV}$, and Ref.~\cite{chen2023anisotropic} $\SI{0.3}{meV}$. Experimental investigations find values in similar ranges~\cite{arregi2023large,Yun2023}.
The systematic dependence of the IL-DMI on the atomic positions of the spacer atoms and on the strength of the SOC, particularly for lighter metals as spacer, remain theoretically unexplored up to now. 

To fill this gap of knowledge, we perform an in-depth theoretical investigation of the IL- and IF-DMI for Co/Me/Co trilayers with Me=Cu, Ag, Pt, Au in hexagonal close-packed (hcp) stacking using fully relativistic first-principles calculations based on the screened Korringa-Kohn-Rostoker method~\cite{szunyogh1994selfconsisent,zabloudil2005electron}. To study the effect of the lattice geometry on the DMI, we apply translations to the atoms of the non-magnetic layer from their symmetric hcp positions. We demonstrate that not only the SOC in the non-magnetic layer but also the SOC in the magnetic layers significantly contributes to the IL-DMI. This fact leads to sizable IL-DMI in Co/Ag/Co trilayers for certain geometries. Interestingly, we find that the material of the spacer plays an important role for the strength of the IL-DMI through band-filling effects even without considering the atomic SOC in the spacer. Therefore, we conclude that the IL-DMI in such trilayers is a complex phenomenon that is closely associated with specific factors such as the charge distribution at the interface. Most of our findings can be qualitatively well explained by adapting the analytical formula for the DMI proposed by Lévy and Fert~\cite{levy1981anisotropy}. Our results are expected to guide future experiments in the search of materials with a large IL-DMI coupling and further the theoretical understanding of magnetic multilayers. Additionally, this work might be a first step towards the understanding of the IL-DMI in disordered systems.

This paper is structured as follows. In Sec.~\ref{sec:method}, we introduce the magnetic interactions obtained from the \textit{ab initio} calculations, explain how we fit the data to a simple macrospin model, show that specific symmetries need to be broken for a non-zero IL-DMI, describe the Lévy-Fert model, and clarify the details of the first-principles method used. In Sec.~\ref{sec:results}, we show the IL-DMI vectors and exchange coupling for different lateral positions of the non-magnetic spacer, verify the fit to the macrospin model by investigating the angle between the magnetizations in the two layers, and present the results for different non-magnetic materials. Finally, in Sec.~\ref{sec:conclusion}, we conclude by summarizing the results and giving an outlook towards possible future works.

\section{\label{sec:method}Methods}
\label{sec:method}
\begin{figure}
    \includegraphics[width=0.99\linewidth]{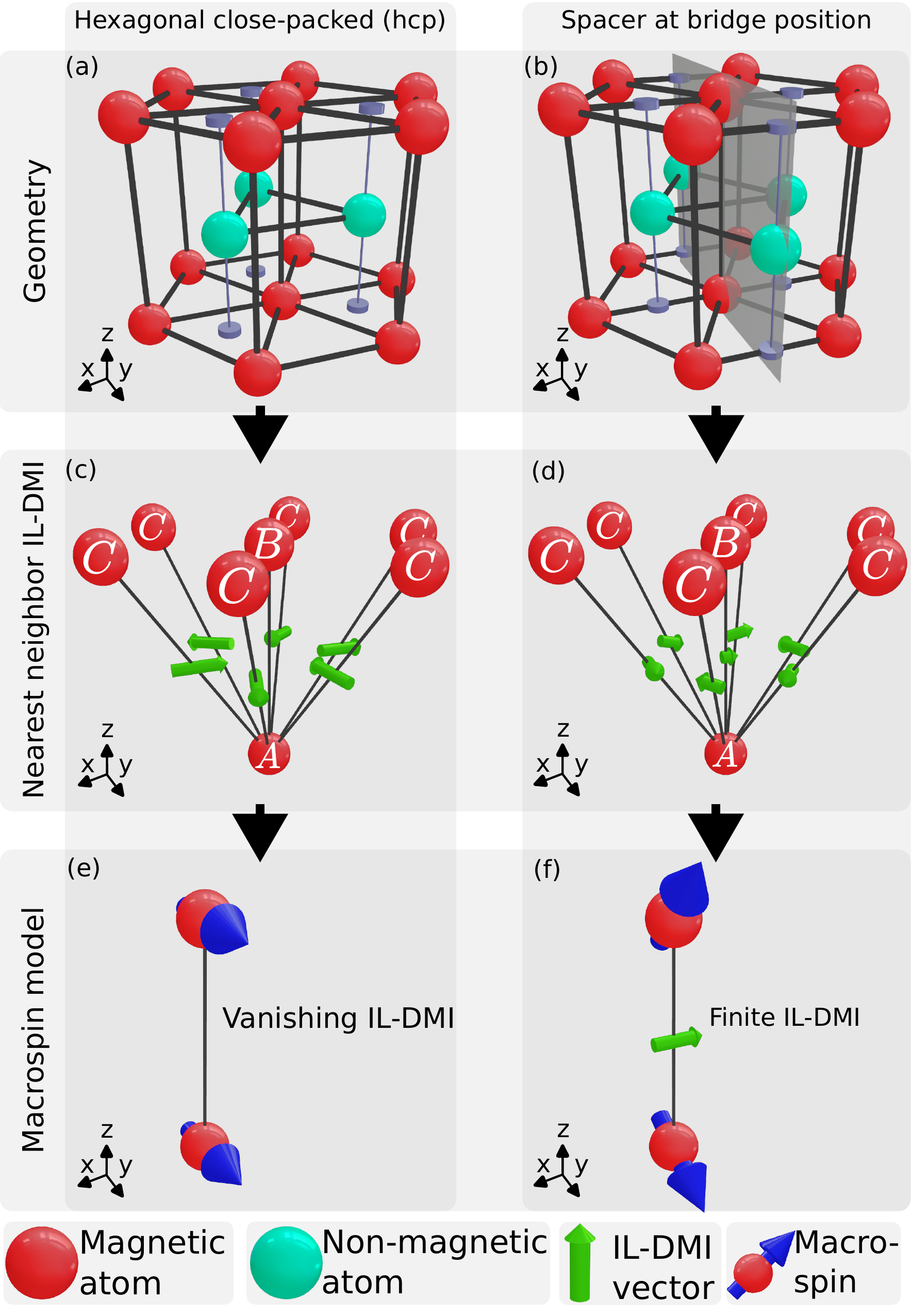}
    \caption{
    Sketch of the atomic positions in the investigated system with atomic nearest-neighbor DMI and resulting IL-DMI in the macrospin model. 
    Based on the known structure of bulk Co, we assumed an hcp structure. (a) and (b) visualize the 
    top bulk layer and the surface magnetic layer with the non-magnetic spacer in (a) hcp stacking or (b) at the bridge position, respectively. (c) and (d) show the
    DMI vectors between atoms in the different magnetic layers obtained from \textit{ab initio} calculations.
    These interactions are fitted to a macrospin model.
    The ground state and the IL-DMI vector for the macrospin model are shown in (e) and (f). The angle between the spins in (f) is exaggerated for illustrative purposes.
    }
    \label{fig:modelSyst}
\end{figure}
\subsection{Crystal symmetry and IL-DMI}
\label{sec:MoriaRules}
The DMI between spin vectors $\vec{S}_{A}$ and $\vec{S}_{B}$ is given by the energy expression $\vec{D}\left(\vec{S}_{A}\times\vec{S}_{B}\right)$. The lattice symmetries of a system determine the direction of the DMI vector $\vec{D}$. The specifics are summarized within the five rules by Moriya~\cite{moriya1960anisotropic}. Denoting the positions of the two magnetic atoms with $A$ and $B$, the straight line connecting them with $AB$, and the center between $A$ and $B$ with $M$, the rules are:
\begin{enumerate}
    \item $M$ is a center of inversion $\Rightarrow \vec{D}=\vec{0}$.
    \item $M$ is in a mirror plane that is perpendicular to $AB$ $\Rightarrow \vec{D} \perp AB$.
    \item $AB$ is in a mirror plane $\Rightarrow \vec{D}\perp \text{mirror plane}$.
    \item A two-fold rotational axis perpendicular to $AB$ passes through $M$ \\$\Rightarrow \vec{D} \perp \text{two-fold rotational axis}$.
    \item An $n$-fold ($n \geq 2$) rotational axis exists along $AB$ \\$\Rightarrow \vec{D} \parallel AB$.
\end{enumerate}

The structure investigated in this work is illustrated in Fig.~\ref{fig:modelSyst}. It consists of a semi-infinite Co bulk in hcp stacking (only the top layer is shown), a single atomic layer of a non-magnetic spacer with varying lateral position, and a single atomic layer of Co directly on top of the atoms below the spacer layer continuing the hcp stacking. 
Consider the case where $B$ and $C$ are nearest neighbors in the same layer, giving rise to IF-DMI. The inversion symmetry is broken by the surface, and the first rule does not apply. For a finite hcp structure with a surface, the second rule applies, resulting in a DMI vector that is perpendicular to the atomic bond. It is important to note that the symmetry breaking at the surface is enough to allow for a non-vanishing IF-DMI even if the spacer layer continues the hcp stacking in Fig.~\ref{fig:modelSyst}(a). 

This changes when looking at the IL-DMI. We now consider $B$ to be in the surface layer and $A$ to be in the top bulk magnetic layer directly below $B$. 
The inversion symmetry is broken because $A$ is in the surface and $B$ is in the bulk. Thus, the first Moriya rule is not satisfied. However, for perfect hcp stacking in Fig.~\ref{fig:modelSyst}(c), the third and fifth rules apply. The $yz$ plane is a mirror plane (see Fig.~\ref{fig:modelSyst}(b)), hence the IL-DMI vector has to point into the $x$ direction. On the other hand, there is a 3-fold rotational symmetry around the $z$ direction. Following the fifth rule, the IL-DMI vector also has to point along the $z$ direction. These two requirements cannot be satisfied simultaneously, thus $\vec{D}=\vec{0}$.
To obtain a nonzero IL-DMI, further symmetries have to be broken. Experimentally, this can be achieved by introducing a gradient in the layer thicknesses~\cite{han2019long}. Here, we investigate the effect of breaking the rotational symmetry by translating atoms of the non-magnetic spacer layer in the $y$ direction, since this is more easily implemented in first-principles calculations. In this case, the mirror symmetry on the $yz$ plane, shown in Fig.~\ref{fig:modelSyst}(b), forces the DMI vector to be along the $x$ direction (Fig.~\ref{fig:modelSyst}(d)), and no further symmetries apply.

Finally, consider $A$ and $C$ in different layers but not directly on top of each other. None of the symmetry rules apply for this pair, and the DMI vector may point along an arbitrary direction. However, the DMI vectors for the six atoms denoted by $C$ are related by symmetry in hcp stacking, as illustrated in Fig.~\ref{fig:modelSyst}(c). Taken together, these interactions give rise to a modulation when moving along the $x$ or $y$ direction, which is functionally the same as for IF-DMI even though the atoms are located at different vertical positions. The sum of these six vectors vanishes, meaning that they do not give rise to an effective IL-DMI on the level of the unit cell \cite{vedmedenko2019interlayer}, as shown in Fig.~\ref{fig:modelSyst}(e). The direction of the macrospin IL-DMI vector follows the symmetry rules for the atoms directly on top of each other. We explain the calculation method for the atomic and macrospin DMI vectors in Secs.~\ref{sec:atomic} and \ref{sec:macrospin}.

\subsection{Magnetic interactions\label{sec:atomic}}
To extract the atomic magnetic interactions, we map the electronic structure of the magnetic system to the following classical atomistic Heisenberg model:
\begin{equation}
    \mathcal{H}=-\frac{1}{2} \sum _{i,j} \vec{S}^\top _i J_{ij} \vec{S}_j - \sum_i \vec{S}^\top _i K_i \Vec{S}_i.
    \label{eq:fullHamiltonian}
\end{equation}
Each lattice site $i$ has a three-dimensional unit vector $\vec{S}_i$ denoting the spin direction. The $3\times 3$ exchange tensor $J_{ij}$ includes all two-spin interactions between the sites $i$ and $j$. The matrix $K_i$ describes the on-site anisotropy tensor. 
For the system in Fig.~\ref{fig:modelSyst}, we only consider sites in the magnetic layers because the magnetic moments in the spacer layer are induced ones for which the Heisenberg model is not appropriate. We determine $J_{ij}$ and $K_i$ from the relativistic generalization~\cite{Udvardi2003} of the torque method~\cite{liechtenstein1987local,tourqueReview2023}, which directly yields interactions between selected pairs of sites without requiring a fitting procedure. Therefore, in the model we only consider sites in the surface layer and the top bulk layer. Inside one layer, we consider all neighbors up to a $5a$ distance, where $a=\SI{2.51}{\angstrom}$ is the in-plane lattice constant of bulk hcp Co. For the hexagonal lattice, we have 90 neighbors for each site. Between the layers, we have the same number of neighbors with the additional interaction between the two Co atoms that are right above each other. Hence, we consider 181 neighbors in total for each site. Further details on the first-principles calculations are given in Appendix~\ref{sec:abinitio}.

To find the magnetic ground state of Eq.~\eqref{eq:fullHamiltonian}, we numerically solve the Landau-Lifshitz-Gilbert equation~\cite{gilbert2004phenomenological},
\begin{equation}
    \dot{\vec{S}}_i = - \frac{\gamma}{1+\alpha^2} \vec{S}_i \times \left( \vec{B}^{\text{eff}}_i + \alpha \vec{S}_i \times \vec{B}^{\text{eff}}_i \right).
    \label{eq:LLG}
\end{equation}
Here, $\gamma$ is the gyromagnetic ratio of an electron, and $\alpha$ is the Gilbert damping which we set to $\alpha=1$ to achieve fast relaxation to equilibrium. The effective magnetic field is determined by the Hamiltonian in the following way:
\begin{equation}
    \vec{B}^{\text{eff}}_i = - \frac{1}{\mu_{i}} \frac{\partial \mathcal{H}}{\partial \vec{S}_i}= \frac{1}{\mu_{i}} \sum_{j} J_{ij} \vec{S}_j + \frac{1}{\mu_{i}} K_i \vec{S}_i.
\end{equation}
where $\mu _i$ is the magnetic moment site $i$. The sum $\sum_j$ runs over all neighbors of $i$. We solve Eq.~\eqref{eq:LLG} using Heun's method by starting from a random initial configuration. We simulate the system till it convergences to an equilibrium state. 
We simulate a total of $64\times 64 \times 2$ spins with periodic boundary conditions inside the layers. We find a single ferromagnetic domain in both layers due to the strong ferromagnetic coupling between the Co atoms inside the layers. The angle between the magnetizations in the two layers allows mapping to a macrospin model discussed in Sec.~\ref{sec:macrospin}.

\subsection{Macrospin model\label{sec:macrospin}}
We want to map the Hamiltonian in Eq.~\eqref{eq:fullHamiltonian} to a simple macrospin model to study the effect of the IL-DMI on the angle between the two ferromagnetic layers. In this model, we describe a given ferromagnetic layer by a single macrospin. 
The Hamiltonian for this model is
\begin{equation}
    H=-J\vec{S}_1\cdot \vec{S}_2 + \vec{D} \cdot \left( \vec{S}_1 \times \vec{S}_2 \right).
    \label{eq:MacroSpin}
\end{equation}
$J$ is the interlayer exchange coupling and $\vec{D}$ is the IL-DMI vector between the two macrospins $\vec{S}_1$ and $\vec{S}_2$. We can rewrite the Hamiltonian of Eq.~\eqref{eq:fullHamiltonian} by introducing isotropic Heisenberg exchange coefficients $\mathcal{J}_{ij}$ and DMI vectors $\vec{\mathcal{D}}_{ij}$~\cite{conte2022coexistence} based on the exchange tensors $J_{ij}$ as follows,
\begin{equation}
    \mathcal{J}_{ij} = \frac{1}{3} \sum_{k\in \{ x,y,z \}} J^{kk}_{ij}, \quad \mathcal{D}^{k}_{ij} = -\frac{1}{2} \sum _{\substack{l \in \{x,y,z\} \\ m \in \{x,y,z\}}}\varepsilon^{klm} J_{ij}^{lm},
\end{equation}
where $\varepsilon^{klm}$ is the Levi-Civita symbol. 
We can write the full Hamiltonian in Eq.~\eqref{eq:fullHamiltonian} using these notations as
\begin{align}
    \mathcal{H}=-\frac{1}{2} \sum _{i,j} \mathcal{J}_{ij} \vec{S}_i \cdot \vec{S}_j &+\frac{1}{2}\sum _{i,j} \vec{\mathcal{D}}_{ij} \cdot (\vec{S}_i \times \vec{S}_j)\nonumber \\ &-\frac{1}{2}\sum_{i,j} \vec{S}_i^{\top} K_{ij} \vec{S}_j.
    \label{eq:SymmetricHamiltronian}
\end{align}
The anisotropy tensor $K_{ij}$ includes the on-site anisotropies $K_{ii}=2K_{i}$, and the symmetric and traceless part of $J_{ij}$.
To fit the full system to the macrospin model, we set $\vec{S}_i=\vec{S}_1$ if the site $i$ is in the first layer, and we set $\vec{S}_i=\vec{S}_2$ if $i$ is in the second layer. Now, by neglecting the anisotropy terms and the constant energy terms from the interfacial interactions, we obtain that $\mathcal{H}$ in Eq.~\eqref{eq:SymmetricHamiltronian} is of the same form as $H$ in Eq.~\eqref{eq:MacroSpin}, where the correspondence between the coefficients is
\begin{equation}
    J = \sum _{i} \mathcal{J}_{0i}, \quad \text{and} \quad \vec{D} = \sum _{i} \vec{\mathcal{D}}_{0i}.\label{eq:calcJD}
\end{equation}
The sums $\sum_i$ run over the neighbors of the spin with index $j=0$, where we assume that this spin is in the first Co layer and all spins $i$ are in the second Co layer. 
We show the validity of the fit to the simple macrospin model by calculating the ground state of the full interacting Hamiltonian of Eq.~\eqref{eq:fullHamiltonian} from spin-dynamics simulations and then determine the angle between the spins in the two Co layers $\alpha$. 
This value is then compared to the macrospin model where the equilibrium spin directions can be calculated analytically. Here, the two spins are perpendicular to the DM vector $\vec{D}$ with a finite angle between them, which is given by
\begin{equation}
    \alpha  =  \text{atan2}\left( D,J\right).
\label{eq:AnalyticalAngle}
\end{equation}
A derivation of Eq.~\eqref{eq:AnalyticalAngle} is given in Appendix~\ref{sec:macrospinangle}.

\subsection{Lévy-Fert model}
We compare the fitted macrospin parameters to the ones obtained from the Lévy-Fert (LF) model. 
In Ref.~\cite{fert1980role,levy1981anisotropy} a formula for the DM vector between two magnetic sites mediated by a third non-magnetic impurity was derived. 
It was assumed that both the magnetic atoms and the impurity are embedded in low concentrations in a non-magnetic material described by a free-electron dispersion. The magnetic atoms interact with the bulk through an exchange coupling between the localized spins and the conduction electrons, while the SOC term on the non-magnetic impurity causes a scattering of the electrons. 
Using third-order perturbation theory, they were able to obtain an expression for the DMI reading
\begin{align}
    \vec{D}_{\text{LF}}=-V_1\frac{\sin \left[ k_F \left( R_{A}+R_{B}+R_{AB} \right) + \phi \right]  }{{R_{A}}^3 {R_{B}}^3 R_{AB}} \nonumber \\ \times \left[ \vec{R}_{A} \cdot \vec{R}_{B} \left( \vec{R}_{A} \times \vec{R}_{B} \right) \right],
    \label{eq:LF_DMI1}
\end{align}
where $R_{A}$ and $R_{B}$ are the distances of the magnetic atoms from the non-magnetic scatterer, $R_{AB}$ is the distance between the magnetic atoms, $k_{\textrm{F}}$ is the Fermi wave vector of the matrix, $\phi = \frac{\pi}{10}Z_d$ is the phase shift induced by the $d$ orbitals of the non-magnetic impurity on the conduction electrons at the Fermi level, and the prefactor $V_{1}$ includes information on the bulk, the magnetic atoms and the non-magnetic scatterer. Notably, $V_{1}$ is proportional to $\lambda_d \sin \left( \phi \right)$ with $\lambda_d$ being the atomic SOC constant, listed in Tab.~\ref{tab:SOCPara} for all non-magnetic spacers used in this work. The DMI vectors naturally satisfy the symmetry rules in Sec.~\ref{sec:MoriaRules} when summed up over the positions of the non-magnetic impurities necessary to satisfy the given symmetry operation. The anisotropies calculated from the analytical expression of the DMI in Eq.~\eqref{eq:LF_DMI} were demonstrated~\cite{levy1981anisotropy} to correlate well with experiments performed on CuMnT alloys when the material T of the non-magnetic scatterer was varied, although the numerical values were higher by about a factor of 100.

\begin{table}
    \centering
    \begin{tabular}{|c||c|c|c|c|c|}
    \hline
         Element & Co & Cu & Ag & Pt & Au\\
         \hline
        $\lambda_d$& \SI{0.068}{eV}  &\SI{0.10}{eV} &\SI{0.22}{eV} &\SI{0.51}{eV} & \SI{0.60}{eV} \\
        
        Atomic $Z_d$ & 7& 10& 10& 9& 10\\
        $ \sin \left(Z_d \pi/10 \right)$ & 0.81 & 0.0 & 0.0& 0.31 & 0.0\\
        
        $\lambda_d \sin \left(Z_d \pi/10 \right)$ & \SI{0.021}{eV} & \SI{0.0}{eV} & \SI{0.0}{eV} & \SI{0.16}{eV}& \SI{0.0}{eV}\\
         \hline
    \end{tabular}
    \caption{Atomic SOC constant $\lambda_d$ and number of $d$-electrons for an isolated atom for different chemical elements considered in the simulations. We also give values for the sinusoidal factor $\lambda_d \sin \left(Z_d \pi/10 \right)$ which is proportional to the DMI strength in the Levy-Fert model~\cite{fert1980role,levy1981anisotropy}. Values are taken from Ref.~\cite{griffith1961theory}. The conversion rate to eV is adopted from Ref.~\cite{levy1981anisotropy}.}
    \label{tab:SOCPara}
\end{table}

We adapt this formula for the systems investigated in our work to calculate the strength of the IL-DMI in the macrospin model. We will position the two magnetic atoms in the bottom layer and the top layer directly above each other. The non-magnetic layer is modeled by a two-dimensional triangular lattice. We define
\begin{equation}
    \vec{r}_A = \begin{pmatrix} 0\\0\\0 \end{pmatrix},\quad \vec{r}_B = \begin{pmatrix} 0\\0\\c \end{pmatrix}, \quad \vec{r}_C^{kl} = \begin{pmatrix} \Delta x + k\cdot a + l \cdot \frac{a}{2}\\ \Delta y + l\cdot \frac{\sqrt{3}}{2} a \\ \frac{c}{2} \end{pmatrix},
\end{equation}
where we introduce the variables $\Delta x$ and $\Delta y$ that shift the entire triangular lattice in $x$ or $y$ direction, while $k$ and $l$ index the positions of the non-magnetic atoms in lattice vector units inside the triangular lattice. We define $\vec{R}_{A/B}^{kl}=\vec{r}
_{A/B}-\vec{r}^{kl}_C$ as new coordinates following the convention in Eq.~\eqref{eq:LF_DMI1} where the distances are measured from the impurity. 
We can now write the DMI vector obtained from the LF model mediated by a selected non-magnetic impurity $kl$ as
\begin{align}
    \vec{D}_{\text{LF}}^{kl}=-V_1\frac{\sin \left[ k_F \left( R_A^{kl}+R_B^{kl}+c \right) + \phi \right]  }{\left(R_A^{kl}\right)^3 \left(R_B^{kl}\right)^3 c} \nonumber \\ \times \left[ \vec{R}_A^{kl} \cdot \vec{R}_B^{kl} \left( \vec{R}_A^{kl} \times \vec{R}_B^{kl} \right) \right].
    \label{eq:LF_DMI}
\end{align}

The DMI vector $\vec{D}$ of the macrospin model Eq.~\eqref{eq:MacroSpin} corresponds to the sum over all lattice sites in the non-magnetic layer in this model, 
\begin{equation}
    \vec{D}_{\text{LF}}^{\textrm{tot}} = \sum _{kl} \vec{D}^{kl}_{\text{LF}}.
    \label{eq:LF_DMI_tot}
\end{equation}
Some approximations are necessary when adapting the model to the present case, since we are investigating a bulk magnet with a non-magnetic spacer instead of a spin glass with a low concentration of impurities. Since the $d$ band is fully occupied in atomic Cu, Ag, and Au ($Z_{d}=10$), we get $\phi=\pi$. We use this value in the phase shift of the sine function, but not in the prefactor $V_{1}$ which is proportional to $\sin \left( \phi \right)$ and would consequently vanish. For $k_F$, we substitute the value obtained for bulk Ag ($k_F=\SI{1.20}{\angstrom^{-1}}$)~\cite{kittel2018introduction} since there is no non-magnetic bulk material here. The Fermi surface for Co, apart from being spin split, does not resemble a sphere~\cite{batallan1974cobaltfermisurface}, meaning that a definition of a single Fermi wave vector for Co is not feasible. We found that the model is not very sensitive to $k_F$, any typical values of $k_F \approx \SI{1}{\angstrom^{-1}}$ for metals produce qualitatively similar results. We use $V_{1}$ as a fitting parameter since we expect no quantitative agreement using the value from Ref.~\cite{levy1981anisotropy}. We focus on investigating how the predicted interaction strength depends on the displacement of the non-magnetic spacer $\Delta x$ and $\Delta y$, on the SOC in the non-magnetic spacer and on the $d$-band filling $Z_{d}$, since these can be compared to the first-principles calculations. In case of a favorable comparison, we expect that the conclusions drawn from the analytic formula in Eq.~\eqref{eq:LF_DMI} may be used to predict a way to optimize material composition for maximizing the IL-DMI.

The interlayer exchange interaction $J$ in Eq.~\eqref{eq:MacroSpin} may be compared to similar analytical formulae derived from perturbation theory. The leading term is the RKKY interaction,
\begin{equation}
    J_{\text{RKKY}}=V_0 \frac{\cos \left( 2 k_F R_{AB} \right) }{R_{AB}^{3}},
    \label{eq:LF_RKKY_Original}
\end{equation}
which only includes contributions from the perturbation at the two magnetic sites, since the presence of a non-magnetic scatterer or SOC is not necessary for its emergence. Unfortunately, this also means that it is independent of the displacement of the spacer layer which does not agree with our first-principles calculations. It is mentioned in Ref.~\cite{levy1981anisotropy} that a correction to the RKKY interaction appears if the scattering off the non-magnetic site is also taken into account. We provide a derivation of this term in Appendix~\ref{sec:RKKY}, which may be approximated as
\begin{align}
    J^{kl}_{\text{RKKY}}=&V'_0 \frac{\sin \left[ k_F \left( R^{kl}_{A}+R^{kl}_{B}+c \right)+\phi' \right] }{\left(R^{kl}_{A}\right)^{3}\left(R^{kl}_{B}\right)^{3}c} \nonumber \\
    &\times \left[3\left(\vec{R}^{kl}_{A}\cdot\vec{R}^{kl}_{B}\right)^{2}-\left(R^{kl}_{A}\right)^{2}\left(R^{kl}_{B}\right)^{2}\right].
    \label{eq:RKKYchanged}
\end{align}
When comparing to the macrospin parameter determined from first-principles calculations, we sum over the non-magnetic sites $kl$,
\begin{align}
J_{\text{RKKY}}^{\textrm{tot}}=\sum_{kl}J^{kl}_{\text{RKKY}},\label{eq:LF_RKKY_total}
\end{align}
and use $V'_{0}$ as a fitting parameter. We also add a constant shift $J_{0}$ to the calculated interaction which includes the contribution coming from the RKKY interaction between the Co atoms in Eq.~\eqref{eq:LF_RKKY_Original}.

\section{\label{sec:results}Results}
\subsection{Lateral shifting of the spacer layer in Co/Ag/Co}
\begin{figure}
    \includegraphics[width=0.99\linewidth]{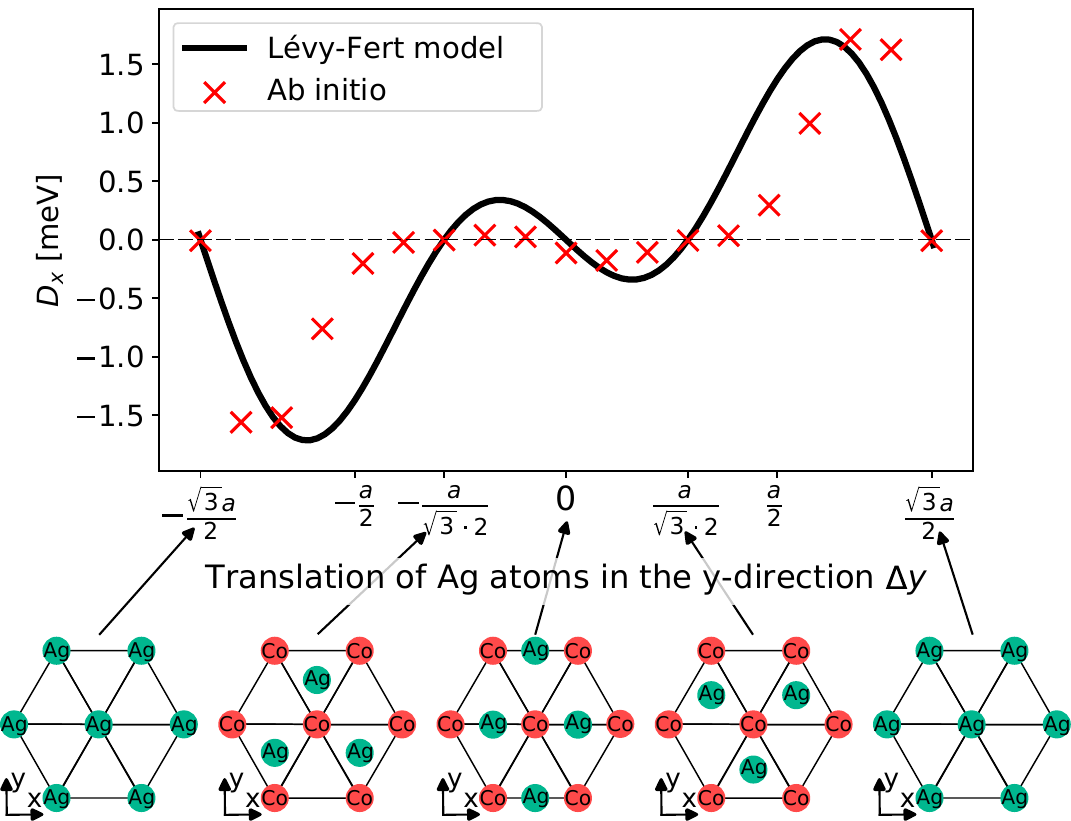}
    \caption{IL-DMI of the macrospin model for different positions of the Ag atoms along the $y$ axis. We compare the strength of the IL-DMI obtained by \textit{ab initio} calculations (red crosses) with the ones predicted by the Lévy-Fert model \eqref{eq:LF_DMI_tot} (black line). The amplitude $V_{1}$ in the Lévy-Fert model was fitted to the \textit{ab initio} data points. The horizontal axis indicates the position of the Ag atoms along the $y$ direction, as shown below the axis. The vertical axis gives the IL-DMI strength for the macrospin model in Eq.~\eqref{eq:MacroSpin}. 
    We only show $D_x$ because the other components are zero.}
    \label{fig:DMIShiftY}
\end{figure}
We start by investigating the effect of a translation of the non-magnetic spacer layer on the parameters of the macrospin model in Eq.~\eqref{eq:MacroSpin}. Initially, we will only look at the Ag spacer, as Co/Ag/Co trilayers were experimentally realized and their IL-DMI measured~\cite{arregi2023large}. 
We shift the Ag atoms along the $y$ direction, calculate the interaction tensors for several positions, and determine the macrospin parameters using Eq.~\eqref{eq:calcJD}. We plot the IL-DMI based on first-principles calculations compared with results of the Lévy-Fert model Eq.~\eqref{eq:LF_DMI_tot} in Fig.~\ref{fig:DMIShiftY}. We denote the displacement of the Ag atoms from the bridge position with $\Delta y$. As discussed in Sec.~\ref{sec:MoriaRules}, the IL-DMI vanishes if the system possesses both threefold rotational symmetry around the $z$ axis and a $yz$ mirror plane. This is satisfied at four points: at $\Delta y=a/(2\sqrt{3})$ and $\Delta y=-a/(2\sqrt{3})$, where hcp and fcc stackings of the spacer layer are achieved, respectively; and for $\Delta y=\pm \sqrt{3}a/2$, where Ag and Co align on top of each other. A shift along the $y$ direction with a different $\Delta y$ value breaks the three-fold rotational symmetry, but the reflection symmetry on the $yz$ plane is preserved. 
Consequently, the DMI vector points in the direction perpendicular to the $yz$ plane, $\vec{D}=\begin{pmatrix} D_x & 0 & 0 \end{pmatrix}$, as discussed in Sec.~\ref{sec:MoriaRules}. Across all values of $\Delta y$, we observe that $D_y$ and $D_z$ vanish for both the \textit{ab initio} and Lévy-Fert model calculations, in agreement with Moriya's rules. Therefore, only the $x$ component of the vector is shown in Fig.~\ref{fig:DMIShiftY}. 
The proportionality constant $V_1$ in Eq.~\eqref{eq:LF_DMI} is set in such a way that the maxima of the model and the \textit{ab initio} ($\max D_x =\SI{1.71}{meV}$) are the same. The overall characteristics of the curve from the simple Lévy-Fert model agree well with the \textit{ab initio} data points. Particularly, the sign changes, the vanishing DMI at $\Delta y=\pm a / (2\sqrt{3})$ and $\Delta y=\pm\sqrt{3}a/2$, and the maximum DMI strength being reached towards the outer limits of the plot are all captured by the Lévy-Fert model. Note that the IL-DMI also vanishes at the bridge position in the Lévy-Fert model. This happens merely due to the fact that only three layers were included in the model calculations, in which case there is a center of inversion between the two Co atoms for this position of the spacer layer, leading to a vanishing DMI. 
In the DFT calculations bulk Co is also considered which breaks this inversion symmetry. However, note that also in the first-principles data there is a sign change in the IL-DMI close to the bridge position of the spacer layer. 
Around the bridge position, the \textit{ab initio} data points are flatter, i.e., closer to zero compared to the curve from the model, and the peaks of maximum DMI strength are closer to $\Delta y=\pm \sqrt{3}a/2$. This is likely due to the limitations of the LF model as similar deviations from DFT results were reported in Refs.~\cite{ga2022dzyaloshinskii,ga2023layer,schweflinghaus2016role}. The experiments on Co/Ag/Co obtain IL-DMI values of about \SI{0.112}{meV}~\cite{arregi2023large}, which is in a similar range compared to our results. To compare the IL-DMI with the IF-DMI by approximating the nearest-neighbor IF-DMI. We find that it lies in the range $-0.4~\text{meV} < D_{\text{IF}} < 0.4~\text{meV}$, making it comparable to the IL-DMI. The details of calculations and the data for the nearest-neighbor IF-DMI are given in Appendix~\ref{sec:appendixIFDMIFit}.
\begin{figure}
    \includegraphics[width=0.99\linewidth]{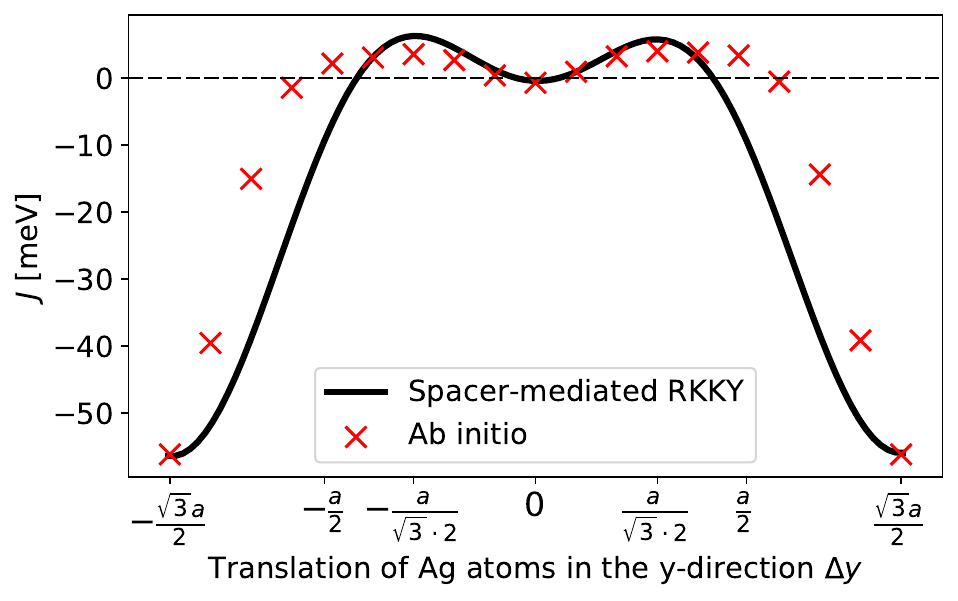}
    \caption{Interlayer exchange interaction of the macrospin model for different positions of the Ag atoms along the $y$ axis. We compare the values of $J$ in Eq.~\eqref{eq:MacroSpin} obtained from \textit{ab initio} calculations (red crosses) with the ones predicted by the spacer-mediated RKKY interaction~\eqref{eq:LF_RKKY_total} (black line). The amplitude $V'_{0}$ and the energy offset $J_0$ are fitted to the \textit{ab initio} data points. The horizontal axis indicates the position of the Ag atoms along the $y$ direction, as depicted in Fig.~\ref{fig:DMIShiftY}.}
    \label{fig:RKKY}
\end{figure}
\begin{figure}
\includegraphics[width=0.99\linewidth]{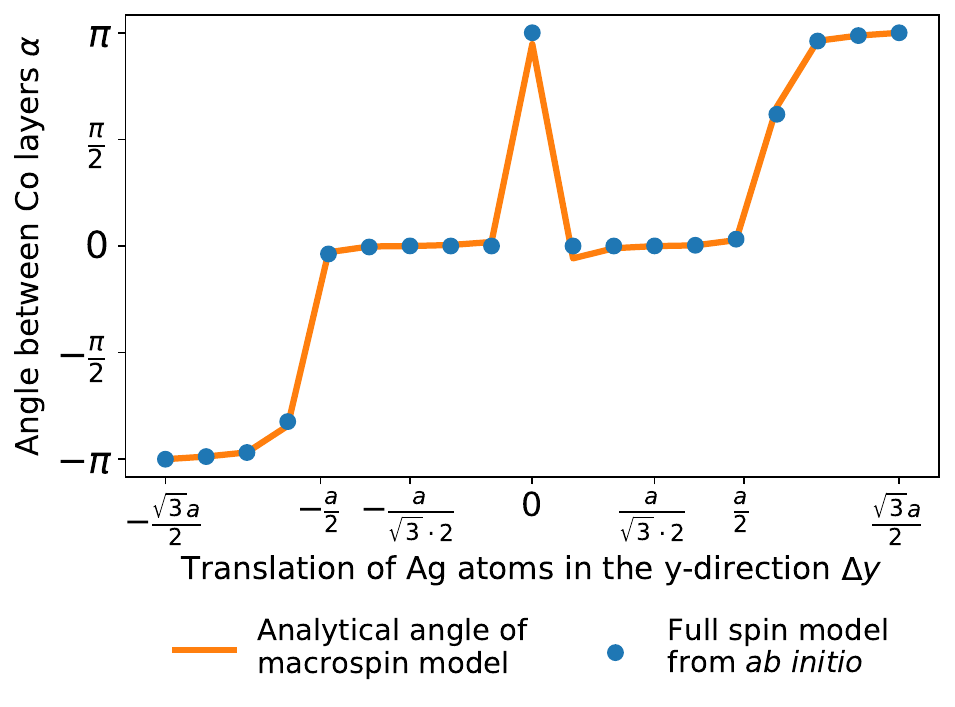}
    \caption{Angle between the two Co layers in the magnetic ground state. The blue dots indicate the values obtained from spin-dynamics simulations using the full Hamiltonian \eqref{eq:fullHamiltonian} including tensorial exchange 
    interactions up to a distance of $5a$. 
    The angle for the macrospin model (orange line) is calculated from Eq.~\eqref{eq:AnalyticalAngle} by taking the values of $\vec{D}$ and $J$ from Fig.~\ref{fig:DMIShiftY}, and Fig.~\ref{fig:RKKY}, respectively. The horizontal axis indicates the position of the Ag 
    atoms along the $y$ direction, as depicted in Fig.~\ref{fig:DMIShiftY}.}
    \label{fig:CoCoAngle}
\end{figure}
\begin{figure*}
    \centering
    \includegraphics[width=0.9\linewidth]{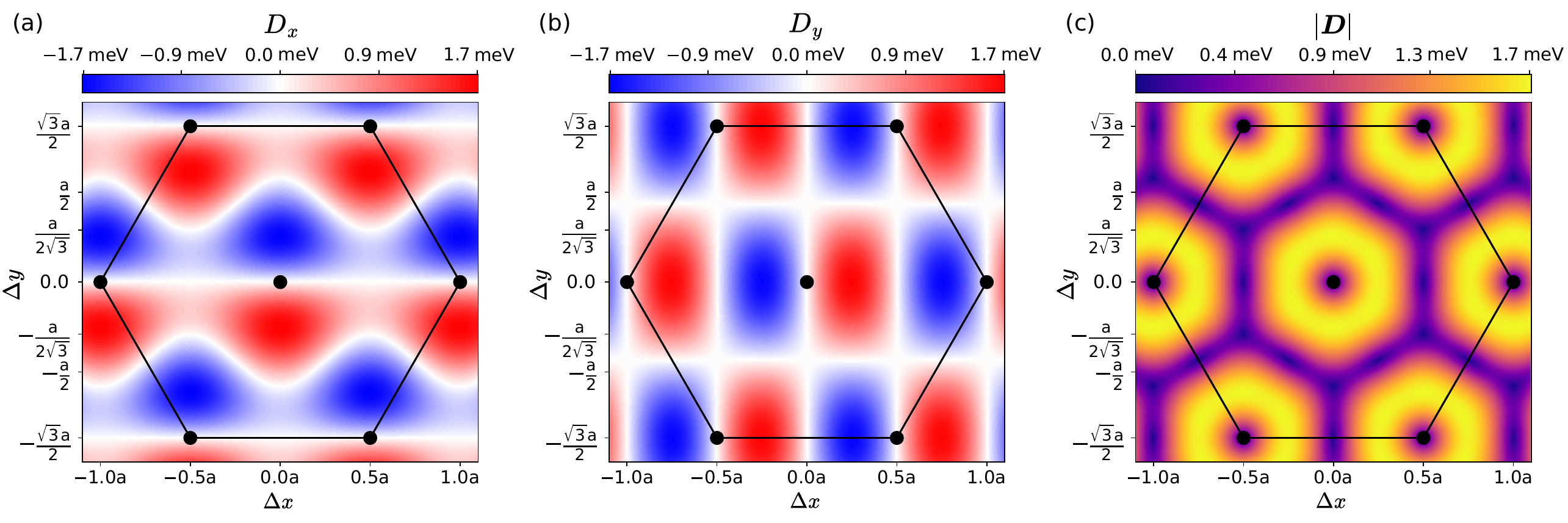}
    \caption{IL-DMI vector obtained from model calculations for different positions of the Ag atoms 
    in the whole unit cell. The vectors are calculated from the Lévy-Fert model Eq.~\eqref{eq:LF_DMI_tot} using the fitting parameters obtained in Fig.~\ref{fig:DMIShiftY} from \textit{ab initio} data along the line $\Delta x=\frac{a}{2}$. 
    Black dots indicate the positions of the Co atoms. (a) shows the $x$ component $D_x$, (b) the $y$ component $D_y$, and (c) the magnitude $|\vec{D}|$ of the DM vector. 
    The colors indicate the IL-DMI strength in the units of meV. 
    }
   \label{fig:unitcelldmi}
\end{figure*}

Additionally to the IL-DMI, we investigate the Heisenberg exchange interaction between the two Co layers in the macrospin model. 
We plot the values for the \textit{ab initio} and model calculations in Fig.~\ref{fig:RKKY}. To determine the proportionality constant $V'_{0}$ in Eq.~\eqref{eq:RKKYchanged} and the offset $J_{0}$, we fit to the \textit{ab initio} data points. Similar to Fig.~\ref{fig:DMIShiftY}, we find that the main characteristics are captured by the model. The local minimum at $\Delta y=0$ and $\Delta y=\pm\sqrt{3}a/2$ are present in both the model and the first-principles data. 
As for the DMI strength, the extrema in the model are at smaller values of $\Delta y$ compared to the \textit{ab initio} data. Note that the interlayer exchange is strongly modulated and even changes sign when changing the position of the spacer layer while keeping the distance between the Co layers fixed. This indicates that the spacer-mediated interaction in Eq.~\eqref{eq:RKKYchanged} has a strong contribution compared to the direct RKKY term in Eq.~\eqref{eq:LF_RKKY_Original}. The sign changes are also captured by Eq.~\eqref{eq:RKKYchanged}, leading to a relatively small fitted offset of $J_0=-\SI{3.80}{meV}$. For the proportionality constant, we obtain $V_0'=\SI{628.3}{meV\cdot\angstrom^3}$.

\subsection{Investigating the angle between ferromagnetic layers}

To demonstrate that the macrospin model accurately describes the ground state, we compare the ground state of a $64 \times 64 \times 2$ spin system obtained from spin-dynamics simulations using the full interaction tensors with the ground state of the corresponding macrospin model. Inside one layer we always obtain a single ferromagnetic domain in the simulations, so we can calculate the angle $\alpha$ between the magnetization directions in the top and bottom layers. We compare this angle to the analytical expression for the angle in the macrospin model Eq.~\eqref{eq:AnalyticalAngle} in Fig.~\ref{fig:CoCoAngle}. The difference between the two methods is caused by the anisotropy terms present in the spin-dynamics simulations but neglected in the macrospin model. However, the two angles largely agree with each other in the figure, indicating that the role of the anisotropy is weak. The angle $\alpha$ is mostly close to either $\alpha = 0$ or $\alpha = \pi$ which correspond to a ferro- ($J>0$) or antiferromagnetic ordering ($J<0$), respectively. These two values are dominant because the amplitude of $J$ is large compared to $D_x$. The peak at $\Delta y=0$ is explained by the sign change in $J$ at $\Delta y=0$, see Fig.~\ref{fig:RKKY}. Around $\Delta y=0$, the two angles deviate from each other. The role of the anisotropy parameters is pronounced in this regime because of the relatively small values of $\vec{D}$ and $J$, see Figs.~\ref{fig:DMIShiftY}~and~\ref{fig:RKKY}.

\subsection{IL-DMI for all possible translations}
Since we found that the Lévy-Fert model reproduces the dependence of the IL-DMI on the position of the atoms in the spacer layer qualitatively well, we use this model instead of the more involved first-principles calculations to investigate the macrospin model parameters for all possible translation vectors $\left( \Delta x,\Delta y \right)$. In Fig.~\ref{fig:unitcelldmi}(a), (b), and (c) we plot the $D_x$, $D_y$, and $|\vec{D}|$ values obtained from the Lévy-Fert model Eq.~\eqref{eq:LF_DMI_tot} using the fitted value of $V_{1}=\SI{58.223}{meV \cdot \angstrom^3}$ from Fig.~\ref{fig:DMIShiftY} over a wider area in two dimensions. The pattern is periodic with the lattice vectors, as indicated by the positions of the Co atoms denoted by black dots. Figure~\ref{fig:DMIShiftY} corresponds to the line profile of Fig.~\ref{fig:unitcelldmi} along $\Delta x=a/2$. Along this line, the $D_y$ component in Fig.~\ref{fig:unitcelldmi}(b) is zero due to the mirror symmetry as discussed in Sec.~\ref{sec:MoriaRules}. The $y$ component in Fig.~\ref{fig:unitcelldmi}(b) follows a rectangular symmetry, while the $x$ component in (a) has a different symmetry with oscillatory characteristic. The absolute value of the DM vector in Fig.~\ref{fig:unitcelldmi}(c) follows the hexagonal symmetry, with rings of strong DMI forming around the positions of the Co atoms. The IL-DMI vanishes ($|\vec{D}|=0$) when the Ag atom is on top of the Co atoms (black dots) and at the fcc and hcp positions (middles of the triangles of black dots) because the threefold rotational symmetry is restored at these points. The IL-DMI also vanishes at the bridge positions (halfway between nearest-neighbor black dots) because the three-layer system used for the model becomes inversion symmetric. While this is not the case in the first-principles calculations, in Fig.~\ref{fig:DMIShiftY} we found that the IL-DMI also vanishes in those calculations somewhere close to the bridge positions. $D_z$ vanishes in the model because the DMI vectors from Eq.~\eqref{eq:LF_DMI} are always perpendicular to the line connecting the two Co atoms which is along the $z$ axis. This feature is not expected from the first-principles calculations when the mirror symmetry on the $yz$ plane is also broken.
\begin{figure}
\includegraphics[width=0.99\linewidth]{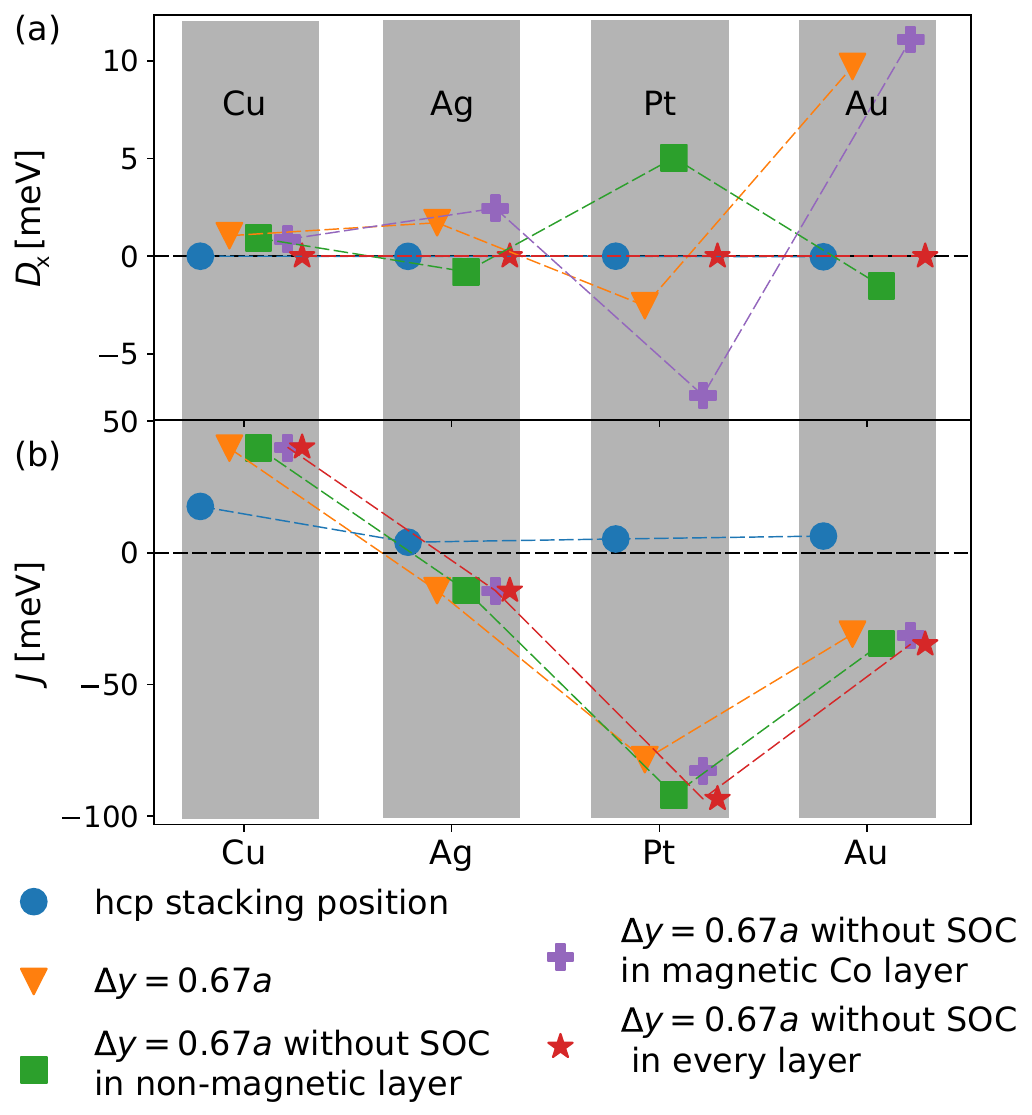}
    \caption{Interlayer interactions for different spacer materials. We obtain the IL-DMI (a) and exchange interactions (b) from the macrospin model Eq.~\eqref{eq:MacroSpin} based on \textit{ab initio} calculations for Cu, Ag, Pt, and Au as the non-magnetic layer. We place the non-magnetic atoms in an hcp stacking position (blue circles), and at the position of maximum DMI for Ag in Fig.~\ref{fig:DMIShiftY}, $\Delta y=0.67a$ (orange triangles). For this latter position, we turn off the SOC in the non-magnetic layer (green squares), in the magnetic Co layers (purple crosses), and in the whole system (red stars). }
    \label{fig:DiffMat}
\end{figure}
\begin{figure}
    \centering
    \includegraphics[width=0.99\linewidth]{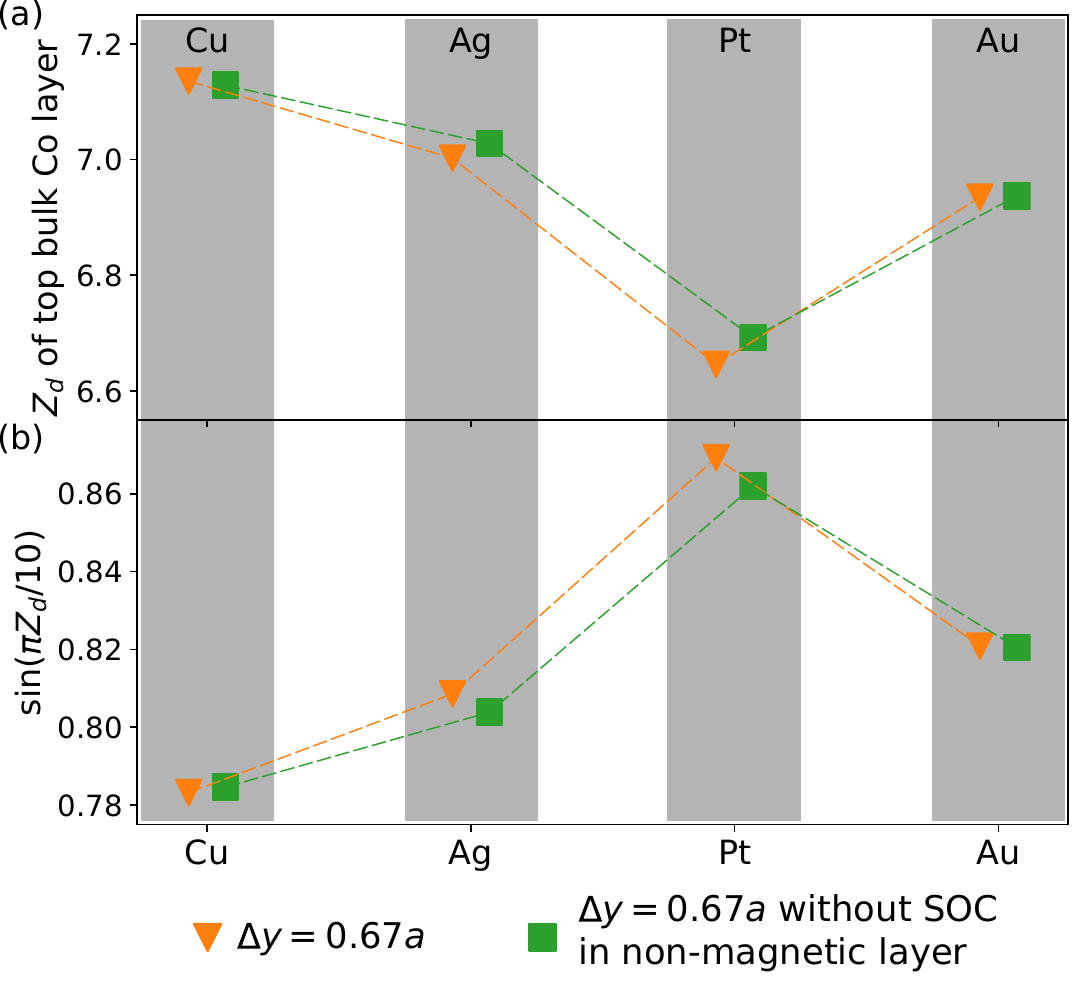}
    \caption{(a) Number of valence $d$ electrons $Z_d$ and (b) prefactor $\sin( (\pi/10) Z_d)$ of the DMI vector in the Lévy-Fert model in the top bulk Co layer for different spacer materials. We chose the geometry with the highest value of DMI for Ag spacer layer $\Delta y=0.67a$ (orange triangles), and repeated the calculations by turning off the SOC in the non-magnetic layer for the same geometry results (green squares).}
    \label{fig:botCoZd}
\end{figure}
\subsection{Material dependencies}
We investigated the dependence of the IL-DMI on the spacer material by replacing Ag with Cu, Pt, and Au. The heavier transition metals like Pt and Au have a higher atomic SOC coupling constant~\cite{griffith1961theory}, hence it is expected that they give rise to a stronger IL-DMI. 
We specifically investigate the elements of group 11 (Cu, Ag, and Au) because the number of valence $d$ electrons appearing in the Lévy-Fert model Eq.~\eqref{eq:LF_DMI1} should be comparable ($Z_d\approx 10$), meaning that these three materials mainly differ in the strength of the SOC. In contrast, Pt and Au have a similar strength of atomic SOC, see Table~\ref{tab:SOCPara}, but Pt has a lower number of valence electrons $Z_d\approx 9$. We calculated the IL-DMI and interlayer exchange interaction in the macrospin model from the interaction tensors determined from first-principles calculations for the symmetric hcp stacking position $\Delta y=a/(2\sqrt{3})$ and the $\Delta y=0.67a$ position. We choose these two geometries because we observe the highest IL-DMI interaction for Ag at $\Delta y=0.67a$ in Fig.~\ref{fig:DMIShiftY}, and a vanishing IL-DMI for $\Delta y=a/(2\sqrt{3})$ because of symmetry reasons. 
Furthermore, we investigate the contribution of the non-magnetic and magnetic atoms to the DMI strength by turning off the SOC selectively in the non-magnetic layer, in the magnetic layers, and in the whole system. The results are shown in Fig.~\ref{fig:DiffMat}(a). In the center position (blue circles) and without any SOC (red stars), the DMI vanishes for all material combinations; the former follows from symmetry and the latter confirms that the SOC is necessary for obtaining DMI in the \textit{ab initio} calculations. At $y=0.67a$, the DMI strength increases in the following order: Cu, Ag, Pt, and Au. When disabling the SOC only in the non-magnetic layer, the DMI is weakened for Cu, Ag, and Au. For Pt, the DMI strength is increased. We also observe a sign change for Ag, Pt, and Au when the SOC in the non-magnetic layer is turned off. In this case the DMI originates from the SOC of the Co atoms, and the resulting DMI acts against the one from the non-magnetic material. The purple crosses in Fig.~\ref{fig:DiffMat} show the resulting interactions when the SOC is turned off in all Co atoms, leaving only the non-magnetic atoms with a non-vanishing SOC constant. In this case, the order of the spacer materials according to the magnitude of the IL-DMI values is again Cu, Ag, Pt, and Au.

By summing up the DMI values $D_x$ when SOC is turned off in the non-magnetic atoms (green squares) and the $D_x$ values when the SOC is turned off in the magnetic Co atoms (purple crosses), one obtains a value close to the case where SOC is included everywhere (orange triangles). This supports that the DMI is linear in the strength of the SOC~\cite{moriya1960anisotropic,fert1980role,levy1981anisotropy}, and the contributions from the magnetic atoms and the non-magnetic atoms are additive. Although the contribution of the SOC on the magnetic atoms to the DMI was not calculated in Ref.~\cite{levy1981anisotropy}, from the similarity of the scattering processes it can be assumed that the main parameters influencing its magnitude are also the atomic SOC $\lambda_{d}$ and the number of valence $d$ electrons $Z_{d}$. Concerning the non-magnetic contributions (purple crosses in Fig.~\ref{fig:DiffMat}), the order of the materials precisely follows the strength of the atomic SOC in Table~\ref{tab:SOCPara}. Note that the IL-DMI induced by Au is stronger than by Pt and reaches an objectively quite high value of around $\SI{10}{meV}$, at least for this specific placement of the non-magnetic spacer. This is remarkable since in the case of IF-DMI in Co, Pt is considered mostly ideal for maximizing its value, while Au typically has a weak effect on it; see, e.g., Ref.~\cite{Simon2018} and references therein. We also provide values for the IF-DMI in Appendix~\ref{sec:appendixIFDMIFit} where we obtain the highest value for Pt. From the data in Fig.~\ref{fig:DiffMat}(a), we can conclude that for Ag and Au, the contribution originating from the magnetic Co is small compared to the overall DMI strength. However, for Cu and Pt the DMI originating from the Co atoms is comparable in strength to the contribution from the non-magnetic atoms. This is less surprising for Cu which has a similar SOC strength as Co, but the strong enhancement of the contribution of Co to the total IL-DMI with a Pt spacer layer is more surprising since the atomic spin-orbit coupling constant $\lambda_d$ is 7.5 times larger for Pt than that of Co~\cite{griffith1961theory}. 

Since the strength of the SOC in Co is not influenced by the material of the spacer layer, we investigate the connection between the IL-DMI attributed to Co and the filling factor $Z_{d}$. The filling factor can be obtained from first-principles calculations by taking the atomically and orbitally resolved DOS, and integrating over the $d$ orbitals up to the Fermi level at the different sites. 
We show this quantity for the different spacer materials in the top bulk Co layer in Fig.~\ref{fig:botCoZd}(a). For bulk Co, one expects $Z_d$ to be close to the atomic filling factor $Z_d \approx 7$~\cite{levy1981anisotropy}. However, due to hybridization with the non-magnetic spacer, $Z_d$ in the Co layer is modified: it is reduced for a Pt spacer layer and increased to a lesser extent for a Cu spacer, while for Ag and Au spacers the filling factor of Co is close to the atomic value. 
The $Z_d$ value controls a phase factor and the DMI strength in the Lévy-Fert model, see Eq.~\eqref{eq:LF_DMI1} and the corresponding discussion. In particular, the prefactor $V_1$ is proportional to $\sin( (\pi/10) Z_d)$. 
In Fig.~\ref{fig:botCoZd}(b) this factor is shown for the top bulk Co layer. The values follow the DMI strength if the SOC is turned off in the non-magnetic spacer (green squares in Fig.~\ref{fig:DiffMat}(a)), indicating qualitative agreement with the generalization of the Lévy-Fert model to SOC scattering off the magnetic atoms when disregarding the geometric factor. 
An explanation based on the filling factors $Z_d$ is not sufficient to completely explain why the contribution of Co to the IL-DMI is close to that of Pt, since Pt and Co both have approximately the same value $Z_d \approx 7$, but different values of the atomic SOC. 
Turning off the SOC in the spacer layer has a weak effect on the band filling. Further analysis of the filling factors $Z_d$ is given in Appendix~\ref{sec:appDOS}.

For the interlayer exchange interaction in Fig.~\ref{fig:DiffMat}(b), we find that it is ferromagnetic but relatively weak for all spacer materials when perfect hcp stacking is followed. Its value is enhanced for $\Delta y=0.67a$ where also strong IL-DMI is found, and it switches sign to antiferromagnetic for Ag, Au, and Pt spacer layers. Turning the SOC off on different atoms has a weak effect on the exchange interaction, since it does not originate from SOC. The IL-DMI has a stronger effect on the angle between the magnetization directions in the layers for Au spacer than for Pt not only because of its higher magnitude, but because the competing interlayer exchange is also weaker.

\section{\label{sec:conclusion}Conclusion}
We investigated the IL-DMI between a surface atomic Co layer and the top atomic layer of bulk Co separated by a non-magnetic spacer layer by combining first-principles calculations, spin model simulations and analytical formulae. For a Ag spacer layer we found a ferromagnetic alignment in each Co layer, and we successfully reduced the magnetic interactions to a macrospin model with a single IL-DMI vector and interlayer exchange constant. We verified the validity of this model by comparing the resulting angles between the ground-state magnetizations in the two Co layers between the full Hamiltonian and the macrospin model. 
We compared the calculated IL-DMI vector and interlayer exchange from the macrospin model with the analytical Lévy-Fert model and its modified version for exchange interactions, and observed qualitatively good agreement as a function of the displacement of the non-magnetic layer. Finally, we investigated the IL-DMI for four different spacer materials and observed an increased DMI strength for heavier elements with a stronger atomic SOC, with the highest value obtained for a Au spacer. We separated the contributions to the IL-DMI from the magnetic and non-magnetic atoms by selectively turning off the SOC in the layers. We found that the contribution of the Co atoms to the IL-DMI partially compensates that of the non-magnetic spacer in the case of Ag, Pt, and Au. This contribution of the magnetic element is particularly enhanced for Pt, which is correlated with the reduced number of valence $d$ electrons in Co for this spacer layer. To the best of our knowledge, this effect of the SOC of the magnetic material has not been reported in the literature before. 
Further investigations are needed to understand the mechanism fully and explain the surprisingly high influence of the relatively weak atomic SOC of Co compared to that of Pt.

We found reasonable agreement between the first-principles calculations and the Lévy-Fert model concerning the dependence of the interactions on the position of the spacer layer, and the enhancement of the IL-DMI for higher atomic SOC strength and for the number of valence electrons $Z_{d}$ being closer to $5$. This indicates that these predictions of the analytical model may be used for exploring larger parameter spaces, but first-principles calculations are still required for determining numerical values of the interactions. Concerning material composition, in the considered geometry we found that the IL-DMI is higher for a Au spacer compared to Pt, while it is known from the literature that the IF-DMI is stronger for Co/Pt than for Co/Au. Even a spacer layer with weaker SOC like Ag may be useful for achieving a large angle between the magnetization directions in the two layers, because the interlayer exchange also strongly varies with the geometry and at some positions it is comparable to the IL-DMI in magnitude. Further studies may uncover material combinations where the contributions from the magnetic and non-magnetic materials to the IL-DMI do not compensate each other, possibly leading to a further enhancement of its magnitude.

In this study, we investigated perfectly ordered crystals, and took into account the symmetry breaking required for the emergence of the IL-DMI by laterally shifting the spacer layer. In experiments the IL-DMI is caused by the growth process introducing a thickness gradient, intermixing at the interfaces and lattice defects which all contribute to the symmetry breaking. Therefore, our work represents a first step towards modeling realistic materials where such disorders also play an important role. 

\section*{Acknowledgments}
E.Y.V. and T.M. acknowledge financial support provided by the Deutsche Forschungsgemeinschaft (DFG) via Project No. 514141286. L.R. gratefully acknowledges funding by the National Research, Development, and Innovation Office (NRDI) of Hungary under Project Nos. K131938 and FK142601, by the Ministry of Culture and Innovation and the National Research, Development and Innovation Office within the Quantum Information National Laboratory of Hungary (Grant No. 2022-2.1.1-NL-2022-00004), and by the Hungarian Academy of Sciences via a J\'{a}nos Bolyai Research Grant (Grant No. BO/00178/23/11).

\appendix
\section{\textit{Ab initio} calculations\label{sec:abinitio}}
The \textit{ab initio} calculations were performed with the screened Korringa–Kohn–Rostoker method (SKKR)~\cite{szunyogh1994selfconsisent,zabloudil2005electron}. 
The fully relativistic nature of the implemented SKKR method enables us to study the effects of SOC in detail; the method enables scaling the strength of the SOC separately in each atomic layer. For more information on this method, we refer to Refs.~\cite{ebert2011calculating,zeller1995theory,papanikolaou2002conceptual}. To study the magnetic properties of two Co layers with a non-magnetic layer between them, we model the system as 10 layers of bulk Co, a monolayer of non-magnetic spacer (Cu, Ag, Pt, or Au), a single Co layer, and four layers of vacuum (empty spheres) between semi-infinite bulk Co and semi-infinite vacuum. 
The two Co layers for which we calculate the interactions and the spacer layer are visualized in Fig.~\ref{fig:modelSyst}(a) and (b). For the stacking, we assume an hcp lattice with the in-plane lattice constant of bulk Co ($a=\SI{2.51}{\angstrom}$). The lattice constant along the $z$ direction is also that of bulk Co ($c=1.63a$) except around the non-magnetic monolayer. Here, we set it to $c=2.06a$ for Ag to preserve the volume of the unit cell of bulk Ag, and similarly to $c=1.68a$ for Cu, $c=1.92a$ for Pt, and $c=2.06a$ for Au. We shift the whole non-magnetic spacer layer along the $y$ direction. This preserves the mirror symmetry on the $yz$ plane. 
For the self-consistent calculations, we used an angular momentum cutoff of $\ell_\text{max}=3$, integrated over 16 energy points on a semicircle contour in the upper complex semiplane, and used 546 $\vec{k}$ points to integrate in the part of the Brillouin zone reduced via the mirror symmetry. The number of $\vec{k}$ points was increased for the calculation of the interactions. 

\section{Determining the angle between the layer magnetizations in the macrospin model}
\label{sec:macrospinangle}
We investigate the Hamiltonian in Eq.~\eqref{eq:MacroSpin}. Expressing the dot and cross products by the angle $\alpha$ between the vectors $\vec{S}_1$ and $\vec{S}_2$ yields
\begin{equation}
    H(\alpha)=-J\cos \left( \alpha \right)+\sin \left( \alpha \right) \vec{D}\cdot \hat{\vec{n}},
\end{equation}
where $\hat{\vec{n}}$ is the normal vector of the plane spanned by $\vec{S}_1$ and $\vec{S}_2$. 
Without loss of generality, we assume that the DMI vector $\vec{D}$ points parallel to the $y$ direction, and set $\hat{\vec{n}}$ along the negative $y$ direction to minimize the energy. Therefore, we replace the DMI vector with a simple scalar value $D$,
\begin{equation}
    H(\alpha)=-J\cos \left( \alpha \right)-D \sin \left( \alpha \right),
\end{equation}
By differentiating we get
\begin{equation}
    \frac{\dd H(\alpha)}{\dd \alpha} = J \sin \left( \alpha \right)- D \cos \left( \alpha \right)=0.
\end{equation}
By rearranging it follows that
\begin{align}
      \tan \alpha &= \frac{D}{J} ,\\
     \alpha &=\text{atan2}\left( D,J \right),
\end{align}
with $\text{atan2}$ resulting in an angle between $-\pi$ and $\pi$ with its sign describing the preferred rotational sense. Assume that the spin $\vec{S}_1$ is pointing along the $z$ direction, $\vec{S}_1=\hat{e}_z$, and a ferromagnetic coupling $J>0$. For $D>0$, the second spin is tilted towards the negative $x$ direction with $0<\alpha <\pi /2$, while for $D>0$, $\vec{S}_{2}$ is tilted towards the positive $x$ direction described by $-\pi/2<\alpha <0$.

\section{Calculation of an effective IF-DMI}
\label{sec:appendixIFDMIFit}
\begin{figure*}
    \centering
    \includegraphics[width=0.99\linewidth]{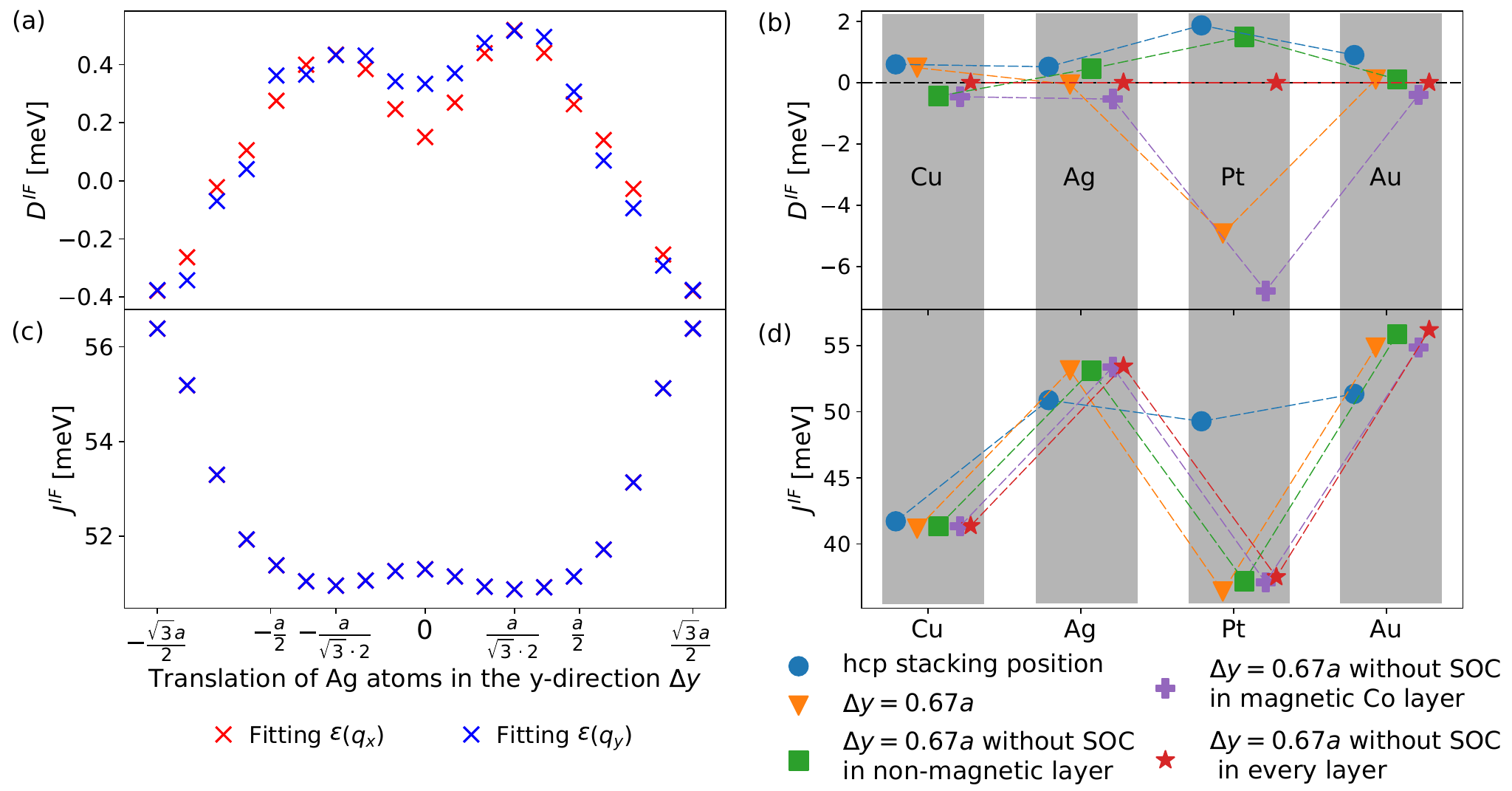}
    \caption{(a) IF-DMI and (c) exchange interactions in an effective nearest-neighbor model Eq.~\eqref{eq:singlelayer} in the surface Co layer for different positions of the non-magnetic Ag spacer layer. The horizontal axis indicates the position of the Ag atoms along the $y$ direction, as visualized in Fig.~\ref{fig:DMIShiftY}. (b) IF-DMI and (d) exchange interactions in the surface Co layer for four different non-magnetic spacer materials (Cu, Ag, Pt, Au). We place the non-magnetic atoms in an hcp stacking position (blue circles), and at the position of maximum IL-DMI for Ag, $\Delta y=0.67a$ (orange triangles). For this position, we turn off the SOC at the non-magnetic layer (green squares), in the magnetic Co layers (purple crosses), and in the whole system (red stars). In (b) and (d) we average over the fitted $J^{IF}$ and $D^{IF}$ from $\epsilon(q_x)$ and $\epsilon(q_y)$ for clarity.}
    \label{fig:IntraInteractionsYShift}
\end{figure*}
The first-principles calculations yield exchange interactions and DM vectors between each pair of magnetic atoms, cf. Eq.~\eqref{eq:SymmetricHamiltronian}. To obtain a simpler measure of the in-plane exchange interaction and the IF-DMI, we approximate the full Hamiltonian restricted to the spins in the surface layer in the low-energy regime by a single-layer triangular lattice with only nearest-neighbor interactions. 
We evaluate the Hamiltonian of Eq.~\eqref{eq:fullHamiltonian} for cycloidal spin spirals $\vec{S}_i (\vec{q})$ with different in-plane wave vectors $\vec{q}$. The spin spiral $\vec{S}_i (\vec{q})$ is given by
\begin{equation}
    \vec{S}_i(\vec{q})=\cos(\vec{q} \cdot \vec{R}_i) \hat{e}_z 
    + \sin( \vec{q} \cdot \vec{R}_i)\hat{q},
    \label{eq:B1}
\end{equation}
where $\hat{e}_z$ is the unit vector pointing into the $z$ direction, $\hat{q}$ is the unit vector pointing into the direction of the in-plane wave vector $\vec{q}$. 

The spins rotating in the plane spanned by $\hat{e}_z$ and $\hat{q}$, called the cycloidal rotation, is preferred by the symmetry restrictions on the IF-DMI vectors in the $C_{3\textrm{v}}$ symmetry class obtained for hcp or fcc stacking.
The Hamiltonian for nearest-neighbor exchange interaction and DMI is given by
\onecolumngrid
\begin{equation}
    H_{NN}^{IF} = -\frac{1}{2} \sum_{\langle i,j \rangle } J^{IF} \vec{S}_i \cdot \vec{S}_j + \frac{1}{2} \sum_{\langle i,j \rangle} \vec{D}_{ij}^{IF} \left( \vec{S}_i \times \vec{S}_j \right).\label{eq:singlelayer}
\end{equation}
We assume that the DMI vectors $\vec{D}_{ij}^{IF}$ are given by $\vec{D}_{ij}=D^{IF} \hat{\vec{R}}_{ij}\times\hat{\vec{e}}_z$.
Now, inserting spin spirals along the $x$ direction and $y$ direction ($\vec{q}=q_{x/y}\hat{\vec{e}}_{x/y}$) from Eq.~\eqref{eq:B1} into the Hamiltonian in Eq.~\eqref{eq:singlelayer}, we obtain
\begin{align}
    \varepsilon (q_x)&=H_{NN}^{IF}\left(\vec{S}_i(q_{x})\right)/N=-J^{IF}  \left[\cos (q_x a)+2\cos\left(q_x\frac{a}{2}\right)\right]-D^{IF} \left[\sin(q_x a)+\sin\left(q_x\frac{a}{2}\right)\right],\\
    \varepsilon (q_y)&=H_{NN}^{IF}\left(\vec{S}_i(q_{x})\right)/N= -J^{IF}\left[ 1+2\cos \left( q_y \frac{\sqrt{3}a}{2}\right)\right]-D^{IF}\frac{\sqrt{3}}{2}\left[2\sin \left( q_y \frac{\sqrt{3}a}{2} \right)\right],
\end{align}
\twocolumngrid
where $N$ is the total number of spins in the system. 
We fit the spin-spiral energies obtained from all magnetic interaction tensors in the surface layer to this expression to obtain the values of $J^{IF}$ and $D^{IF}$. We separate the symmetric and antisymmetric parts of the tensors in Eq.~\eqref{eq:fullHamiltonian} to fit to $J^{IF}$ and $D^{IF}$ separately. We only fit to small $\vec{q}$ values of $\left|q_{x,y}\right|<0.1\frac{2 \pi}{a}$ since we want the model to describe the low-energy regime close to the ferromagnetic ground state.
The fitted interfacial parameters are summarized in Fig.~\ref{fig:IntraInteractionsYShift} for different positions of the Ag spacer layers as well as for different spacer layers and strengths of the SOC. In Fig.~\ref{fig:IntraInteractionsYShift}(a), the IF-DMI has a strong dependence on the position of the non-magnetic layer with some oscillatory features, similarly to the IL-DMI. The key difference is that the IF-DMI does not vanish at any position due to symmetry reasons, since inversion-symmetry breaking by the surface is sufficient for its presence. However, the fitted values along the $q_{x}$ and $q_{y}$ directions are slightly different. This is because assuming $\vec{D}_{ij}=D^{IF} \hat{\vec{R}}_{ij}\times\hat{\vec{e}}_z$ for the in-plane component of the IF-DMI with a single $D^{IF}$ value only follows from the Moriya rules for $C_{3\textrm{v}}$ symmetry, but the DMI vectors for neighbors along the $x$ and $y$ direction become inequivalent as the symmetry is reduced. Note that the fitted values along $q_{x}$ and $q_{y}$ indeed approximately coincide at the high-symmetry points $\Delta y=\pm a/\left(2\sqrt{3}\right),\pm \sqrt{3}a/2$. 
The exchange interaction in Fig.~\ref{fig:IntraInteractionsYShift}(c) has a similar shape as the interlayer exchange interaction in Fig.~\ref{fig:RKKY} but is inverted along the $y$ axis. The coupling inside a layer is always ferromagnetic, and the changes in it are relatively weaker than in the interlayer term. This is probably because scattering processes inside a single Co layer are primarily responsible for the strong ferromagnetic coupling. 
The interfacial interactions for Cu, Ag, Pt, and Au as the non-magnetic spacer layer are displayed in Fig.~\ref{fig:IntraInteractionsYShift}(b) and (d). At the symmetric hcp position, we observe a non-vanishing IF-DMI for all materials which is allowed by symmetry, with Pt having the highest DMI strength. As in Fig.~\ref{fig:DiffMat}, we study the magnetic interactions for a displacement of $\Delta y =0.67a$ along the $y$ direction, for which we observed a maximum in the IL-DMI for a Ag spacer layer. In Fig.~\ref{fig:IntraInteractionsYShift}(a), we see that the IF-DMI is quite small and close to its minimum for a Ag spacer layer at the same displacement. 
At $\Delta y =0.67a$ in Fig.~\ref{fig:IntraInteractionsYShift}(b), we see a decrease in IF-DMI for Cu, Ag, and Au compared to the hcp stacking, but a further increase in Pt. The high value of the IF-DMI at Co/Pt is in agreement with earlier works~\cite{Simon2018}. We also show the data for SOC turned off in the non-magnetic spacer (green squares), in the Co (purple crosses), and in the whole sample (red stars). When turning off the SOC in the whole material, the IF-DMI vanishes in all cases as expected. The exchange interaction in the surface Co layer is always ferromagnetic, it is only influenced relatively strongly by the position of the spacer layer in the case of Pt, and depends weakly on the SOC.

\section{Band-filling effects}
\label{sec:appDOS}
\begin{figure*}
    \centering
    \includegraphics[width=0.99\linewidth]{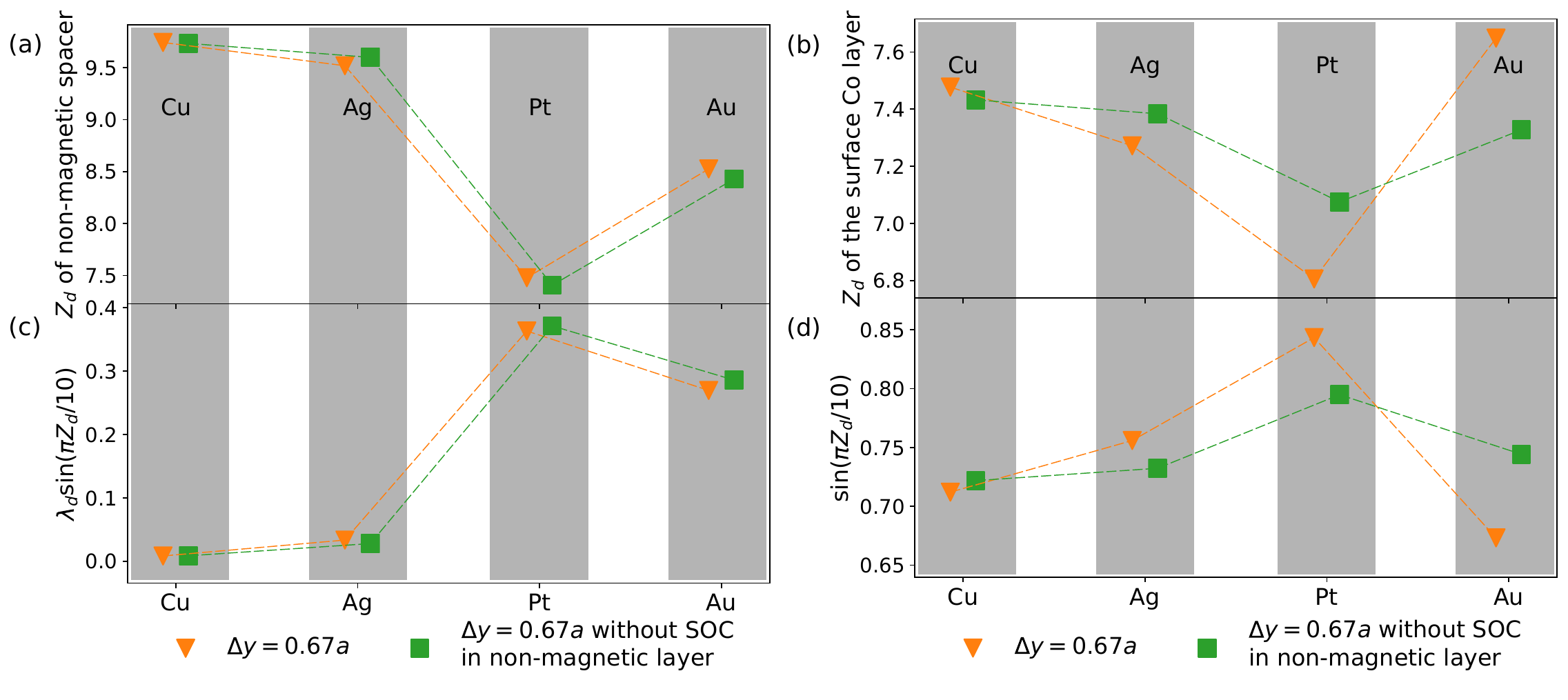}
    \caption{Filling factor $Z_d$ (a) in the non-magnetic spacer layer (b) and the surface Co layer for different spacer materials. (c) and (d) show the factor $V_1=\lambda_d \sin( (\pi/10) Z_d)$ for the same cases, which is proportional to the DMI in the Lévy-Fert model Eq.~\eqref{eq:LF_DMI1}.}
    \label{fig:Zd_rest}
\end{figure*}
To elaborate on the discussion of Figs.~\ref{fig:DiffMat} and \ref{fig:botCoZd}, in Fig.~\ref{fig:Zd_rest} we provide the filling factors $Z_d$ (a) in the non-magnetic spacer layers and (b) in the surface Co layer for different spacer materials. The presented data points in (a) and (b) are for a displacement of the non-magnetic spacer by $\Delta y =0.67a$ along the $y$ direction, at which value the IL-DMI is maximal for a Ag spacer layer. The number of valence $d$ electrons $Z_d$ obtained from the calculations deviates from the atomic value because of hybridization. For Cu, Ag, and Au the atomic filling factor is $Z_d=10$, for Pt it is $Z_d=9$, and for Co it is $Z_d=7$. The prefactor of the DMI vector in the Lévy-Fert model $V_1 \sim \lambda_d \sin( (\pi/10) Z_d)$ related to the filling factor is shown in Fig.~\ref{fig:Zd_rest}(c) and (d). $\lambda_d$ is chosen for each element separately using the specific values given in Tab.~\ref{tab:SOCPara}, but it is not taken into account in panel (d) since all data points are for Co. For Cu, Ag, and Au spacer layers one obtains $\sin( (\pi/10) Z_d)=0$ using the atomic value of $Z_{d}=10$ in the model, which would indicate a vanishing DMI. 
However, Figs.~\ref{fig:Zd_rest}(a) and (c) show that the DMI does not completely vanish here because the actual filling factor is lower than this value. The increasing trend with increasing SOC strength in Fig.~\ref{fig:Zd_rest}(c) is similar to Fig.~\ref{fig:DiffMat}(a) when the SOC in the Co layers is turned off (purple crosses). The prefactor $\lambda_d \sin( (\pi/10) Z_d)$ is higher for Pt than for Au because of the lower filling of Pt, yet the order of the IL-DMI values in Fig.~\ref{fig:DiffMat}(a) is the opposite for these two materials. This may be attributed to the geometric factors not taken into account in the prefactor. Figure~\ref{fig:Zd_rest}(d) may be compared to the IL-DMI values in Fig.~\ref{fig:DiffMat}(a) without SOC in the spacer layer (green squares), or to the same plot for the top bulk Co layer in Fig.~\ref{fig:botCoZd}. The increasing trend in the DMI prefactor from Cu through Ag to Pt is reproduced, but the fact that the predicted DMI prefactor for Au is smaller than for any other spacer layer is in contrast to the calculated IL-DMI values in Fig.~\ref{fig:DiffMat}(a) where the contribution of the Co layer to the DMI is the second highest in Au. However, the top bulk Co layer in Fig.~\ref{fig:botCoZd} and the surface Co layer in Fig.~\ref{fig:Zd_rest}(b) and (d) both influence the contribution of the magnetic atoms to the DMI. Note that the filling factors $Z_{d}$ of both Co layers and that of the Pt layer are below their atomic value for Pt as a spacer. This may be caused in part by partial charge transfer outside these three layers, i.e., to the vacuum layers or the other bulk Co layers, or in part by a hybridization between $s$ and $d$ orbitals transferring charge to the former. Such effects are not possible to incorporate in the Lévy-Fert model, and these limitations have to be kept in mind. 
\onecolumngrid
\section{Derivation of the spacer-mediated RKKY interaction}
\label{sec:RKKY}
Here we calculate the contribution of the non-magnetic atoms to the exchange interaction. We follow the derivation of Ref.~\cite{levy1981anisotropy}, where such a term was predicted but not explicitly calculated. The correction to the ground-state energy of a three-dimensional free Fermi gas in second-order perturbation theory is
\begin{align}
E^{(2)}=\left(\frac{1}{8\pi^{3}}\right)^{2}\int_{k\le k_{\textrm{F}}}\textrm{d}^{3}\boldsymbol{k}\mathcal{P}\int\textrm{d}^{3}\boldsymbol{k}'\sum_{\sigma,\sigma'}\frac{\langle\boldsymbol{k},\sigma|V|\boldsymbol{k}',\sigma'\rangle\langle\boldsymbol{k}',\sigma'|V|\boldsymbol{k},\sigma\rangle}{E_{\boldsymbol{k},\sigma}-E_{\boldsymbol{k}',\sigma'}},\label{eq:D1}
\end{align}
where $E_{\boldsymbol{k},\sigma}=E_{\boldsymbol{k}}=\hbar^{2}k^{2}/(2m)$ is the free-electron dispersion and $\mathcal{P}$ denotes the principal value of the integral. As perturbation we consider two point-like magnetic impurities with spins $\boldsymbol{S}_{a}$ and $\boldsymbol{S}_{b}$ located at positions $\boldsymbol{R}_{a}$ and $\boldsymbol{R}_{b}$,
\begin{align}
V=-\Gamma\delta\left(\boldsymbol{r}-\boldsymbol{R}_{a}\right)\boldsymbol{s}\cdot\boldsymbol{S}_{a}-\Gamma\delta\left(\boldsymbol{r}-\boldsymbol{R}_{b}\right)\boldsymbol{s}\cdot\boldsymbol{S}_{b},\label{eq:D2}
\end{align}
which interact with the spins of the conduction electrons $\boldsymbol{s}$ through the coupling $\Gamma$. Furthermore, we consider that the free-electron eigenstates $|\boldsymbol{k},\sigma\rangle$ include the effect of a perturbation up to first order by a centrosymmetric potential located at a non-magnetic site,
\begin{eqnarray}
|\boldsymbol{k},\sigma\rangle=\textrm{e}^{\textrm{i}\boldsymbol{k}\boldsymbol{r}}-4\pi\textrm{e}^{\textrm{i}\eta_{2}\left(\boldsymbol{k}\right)}\sin\eta_{2}\left(\boldsymbol{k}\right)h_{2}^{(+)}\left(kr\right)\sum_{m=-2}^{2}Y_{2m}^{*}\left(\hat{\boldsymbol{k}}\right)Y_{2m}\left(\hat{\boldsymbol{r}}\right),\label{eq:D3}
\end{eqnarray}
where $\eta_{2}\left(\boldsymbol{k}\right)$ is the phase shift caused by the impurity, $h_{2}^{(+)}$ is the spherical Hankel function of the first kind and $Y_{2m}$ is the spherical harmonic, while the position $\boldsymbol{r}$ is measured from the position of the impurity. The phase shift is expressed in the virtual-bound-state approximation as
\begin{align}
\eta_{2}\left(\boldsymbol{k}\right)=\arctan\frac{\Delta}{E_{r}-E_{\boldsymbol{k}}},\label{eq:D4}
\end{align}
where $E_{r}$ is the position of the resonance and $\Delta$ is its width. Note that we only consider perturbation by the $d$ orbitals of the impurity, leading to the indices $l=2$ in the expressions. We do not consider SOC here since it is not necessary for the correction to the Heisenberg interaction, and the strength of the SOC has a weak effect on this interaction in our first-principles calculations as demonstrated above.

Substituting the unperturbed wave functions $\textrm{e}^{\textrm{i}\boldsymbol{k}\boldsymbol{r}}$ from Eq.~\eqref{eq:D3} into Eq.~\eqref{eq:D1} and performing the integrals results in the RKKY interaction,
\begin{align}
E^{(2)}_{(0)}=\frac{9\pi\Gamma^{2}}{32E_{\textrm{F}}k^{3}_{\textrm{F}}}\frac{\cos\left(2k_{\textrm{F}}R_{ab}\right)}{R^{3}_{ab}}\boldsymbol{S}_{a}\cdot\boldsymbol{S}_{b},\label{eq:D5}
\end{align}
where $k_{\textrm{F}}$ and $E_{\textrm{F}}$ are the Fermi wave vector and velocity, and $R_{ab}=\left|\boldsymbol{R}_{a}-\boldsymbol{R}_{b}\right|$.
When one substitutes the perturbation from Eq.~\eqref{eq:D3} into one of the four wave functions $|\boldsymbol{k},\sigma\rangle$ in Eq.~\eqref{eq:D1} while considering the unperturbed plane waves in the other three wave functions, one gets
\begin{align}
E^{(2)}_{(1)}=&\left(\frac{1}{8\pi^{3}}\right)^{2}\frac{\Gamma^{2}}{N^{2}}\int_{0}^{k_{\textrm{F}}}k^{2}\textrm{d}k\mathcal{P}\int_{0}^{\infty}{k'}^{2}\textrm{d}k'\frac{1}{E_{k}-E_{k'}}\sum_{\sigma,\sigma'}\langle\sigma|\boldsymbol{s}\cdot\boldsymbol{S}_{a}|\sigma'\rangle\langle\sigma'|\boldsymbol{s}\cdot\boldsymbol{S}_{b}|\sigma\rangle\nonumber\\&\times\left[-4\pi\int\textrm{e}^{-\textrm{i}\boldsymbol{k}\left(\boldsymbol{R}_{a}-\boldsymbol{R}_{b}\right)}\textrm{d}\hat{\boldsymbol{k}}\left(\textrm{e}^{-\textrm{i}\eta_{2}\left(k'\right)}\sin\eta_{2}\left(k'\right)h_{2}^{(-)}\left(k'R_{b}\right)\int\textrm{e}^{\textrm{i}\boldsymbol{k}'\boldsymbol{R}_{a}}\sum_{m=-2}^{2}Y_{2m}\left(\hat{\boldsymbol{k}}'\right)Y_{2m}^{*}\left(\hat{\boldsymbol{R}_{b}}\right)\textrm{d}\hat{\boldsymbol{k}}'\right.\right.\nonumber\\&\left.\left.+\textrm{e}^{\textrm{i}\eta_{2}\left(k'\right)}\sin\eta_{2}\left(k'\right)h_{2}^{(+)}\left(k'R_{a}\right)\int\textrm{e}^{-\textrm{i}\boldsymbol{k}'\boldsymbol{R}_{b}}\sum_{m=-2}^{2}Y_{2m}^{*}\left(\hat{\boldsymbol{k}}'\right)Y_{2m}\left(\hat{\boldsymbol{R}_{a}}\right)\textrm{d}\hat{\boldsymbol{k}}'\right)+(a\Leftrightarrow b)\right]+(\boldsymbol{k}\Leftrightarrow\boldsymbol{k}').\label{eq:D6}
\end{align}
Here, $N$ is the atom density resulting from performing the spatial integrals over the delta distributions in Eq.~\eqref{eq:D2}.
The summation over spin indices can be performed separately,
\begin{align}
\sum_{\sigma,\sigma'}\langle\sigma|\boldsymbol{s}\cdot\boldsymbol{S}_{a}|\sigma'\rangle\langle\sigma'|\boldsymbol{s}\cdot\boldsymbol{S}_{b}|\sigma\rangle=\sum_{\alpha,\beta}\frac{1}{4}\sum_{\sigma,\sigma'}\sigma^{\alpha}_{\sigma\sigma'}\sigma^{\alpha}_{\sigma'\sigma}S^{\alpha}_{a}S^{\beta}_{b}=\frac{1}{2}\boldsymbol{S}_{a}\cdot\boldsymbol{S}_{b},\label{eq:D7}
\end{align}
yielding an isotropic exchange interaction.
The integrals over the directions of the wave vectors $\hat{\boldsymbol{k}}$ and $\hat{\boldsymbol{k}}'$ read
\begin{align}
&\int\textrm{e}^{-\textrm{i}\boldsymbol{k}\left(\boldsymbol{R}_{a}-\boldsymbol{R}_{b}\right)}\textrm{d}\hat{\boldsymbol{k}}=\frac{4\pi\sin\left(kR_{ab}\right)}{kR_{ab}},\label{eq:D8}
\\
&\int\textrm{e}^{\textrm{i}\boldsymbol{k}'\boldsymbol{R}_{a}}\sum_{m=-2}^{2}Y_{2m}\left(\hat{\boldsymbol{k}}'\right)Y_{2m}^{*}\left(\hat{\boldsymbol{R}_{b}}\right)\textrm{d}\hat{\boldsymbol{k}}'=-4\pi j_{2}\left(k'R_{a}\right)\sum_{m=-2}^{2}Y_{2m}\left(\hat{\boldsymbol{R}_{a}}\right)Y_{2m}^{*}\left(\hat{\boldsymbol{R}_{b}}\right)=-\frac{5}{2}j_{2}\left(k'R_{a}\right)\left[3\left(\hat{\boldsymbol{R}_{a}}\cdot\hat{\boldsymbol{R}_{b}}\right)^{2}-1\right],\label{eq:D9}
\end{align}
using the properties of the spherical harmonics. Assuming that the magnetic impurities are far away from the non-magnetic one, the spherical Hankel and Bessel functions may be approximated by their asymptotic forms,
\begin{align}
h^{(\pm)}_{2}\left(kr\right)\approx-\frac{\textrm{e}^{\pm\textrm{i}kr}}{kr}, \qquad j_{2}\left(kr\right)\approx-\frac{\sin\left(kr\right)}{kr}.\label{eq:D10}
\end{align}
Collecting the terms in Eq.~\eqref{eq:D6} while exchanging the $a$ and $b$ indices and using trigonometric identities results in
\begin{align}
E^{(2)}_{(1)}=&\frac{5\Gamma^{2}}{8\pi^{4}N^{2}}\frac{m}{\hbar^{2}}\frac{1}{R_{a}R_{b}R_{ab}}\boldsymbol{S}_{a}\cdot\boldsymbol{S}_{b}\left[3\left(\hat{\boldsymbol{R}_{a}}\cdot\hat{\boldsymbol{R}_{b}}\right)^{2}-1\right]\int_{0}^{k_{\textrm{F}}}k\textrm{d}k \sin\left(kR_{ab}\right)\nonumber\\&\times\mathcal{P}\int_{0}^{\infty}\textrm{d}k'\frac{1}{k^{2}-{k'}^{2}}2\left[\sin\eta_{2}\left(k'\right)\sin\left(k'\left(R_{a}+R_{b}\right)+\eta_{2}\left(k'\right)\right)-\sin^{2}\eta_{2}\left(k'\right)\cos\left(k'\left(R_{a}-R_{b}\right)\right)\right]+(\boldsymbol{k}\Leftrightarrow\boldsymbol{k}').\label{eq:D11}
\end{align}
The principal-value integrals may be evaluated using contour integrals in the complex plane as discussed in Ref.~\cite{levy1981anisotropy}. This results in an infinite series in negative powers of the distances $R_{a}+R_{b}$, $\left|R_{a}-R_{b}\right|$, and $R_{ab}$. Keeping only the leading-order terms which decay the slowest with distance gives
\begin{align}
&\mathcal{P}\int_{0}^{\infty}\sin\eta_{2}\left(k'\right)\sin\left(k'\left(R_{a}+R_{b}\right)+\eta_{2}\left(k'\right)\right)\frac{1}{k^{2}-{k'}^{2}}\textrm{d}k'\approx-\frac{\pi}{2k}\sin\eta_{2}\left(k\right)\cos\left(k\left(R_{a}+R_{b}\right)+\eta_{2}\left(k\right)\right),\label{eq:D12}
\\
&\mathcal{P}\int_{0}^{\infty}\sin^{2}\eta_{2}\left(k'\right)\cos\left(k'\left(R_{a}-R_{b}\right)\right)\frac{1}{k^{2}-{k'}^{2}}\textrm{d}k'\approx\frac{\pi}{2k}\sin^{2}\eta_{2}\left(k\right)\sin\left(k\left|R_{a}-R_{b}\right|\right),\label{eq:D13}
\end{align}
and
\begin{align}
\mathcal{P}\int_{0}^{\infty}\sin\left(k'R_{ab}\right)\frac{k'}{k^{2}-{k'}^{2}}\textrm{d}k'\approx-\frac{\pi}{2}\cos\left(kR_{ab}\right),\label{eq:D14}
\end{align}
from the term where $\boldsymbol{k}$ and $\boldsymbol{k}'$ are interchanged. After collecting all the terms and using trigonometric identities, this leads to
\begin{align}
E^{(2)}_{(1)}=&-\frac{5\Gamma^{2}}{8\pi^{3}N^{2}}\frac{m}{\hbar^{2}}\frac{1}{R_{a}R_{b}R_{ab}}\boldsymbol{S}_{a}\cdot\boldsymbol{S}_{b}\left[3\left(\hat{\boldsymbol{R}_{a}}\cdot\hat{\boldsymbol{R}_{b}}\right)^{2}-1\right]\nonumber\\&\times\int_{0}^{k_{\textrm{F}}} \sin\left(k\left(R_{a}+R_{b}+R_{ab}\right)\right)\sin\eta_{2}\left(k\right)\cos\eta_{2}\left(k\right)\nonumber\\&+\left[\cos\left(k\left(R_{a}+R_{b}+R_{ab}\right)\right)-\cos\left(k\left(\left|R_{a}-R_{b}\right|+R_{ab}\right)\right)\right]\sin^{2}\eta_{2}\left(k\right)\textrm{d}k.\label{eq:D15}
\end{align}
Note that $\left|R_{a}-R_{b}\right|$ is the difference between the distances of magnetic atoms $a$ and $b$ measured from the impurity, while $R_{ab}=\left|\boldsymbol{R}_{a}-\boldsymbol{R}_{b}\right|$ is the distance between the magnetic atoms.

Assuming that the virtual bound state is very narrow, the factor $\sin\eta_{2}\left(k\right)$ vanishes for wave vectors away from the Fermi level, and changes very rapidly in its vicinity. Therefore, $k$ may be replaced by $k_{\textrm{F}}$ in the slowly varying trigonometric functions which depend on the relative positions of the atoms in Eq.~\eqref{eq:D15}. Performing the integral in $k$ yields
\begin{align}
&\int_{0}^{k_{\textrm{F}}}\sin\eta_{2}\left(k\right)\cos\eta_{2}\left(k\right)\textrm{d}k=\int_{0}^{k_{\textrm{F}}}\frac{\left(E_{r}-E_{k}\right)\Delta}{\left(E_{r}-E_{k}\right)^{2}+\Delta^{2}}\textrm{d}k=\frac{2m\Delta}{\hbar^{2}}\int_{0}^{k_{\textrm{F}}}\frac{\left(k^{2}_{r}-k^{2}\right)}{\left(k^{2}_{r}-k^{2}\right)^{2}+k_{\Delta}^{4}}\textrm{d}k\nonumber\\&=-\frac{2m\Delta}{\hbar^{2}}\left[\frac{\textrm{e}^{\textrm{i}\frac{\pi}{4}}\arctan\left(\frac{\textrm{e}^{\textrm{i}\frac{\pi}{4}}k}{\sqrt{k^{2}_{\Delta}-\textrm{i}k^{2}_{r}}}\right)}{2\sqrt{k^{2}_{\Delta}-\textrm{i}k^{2}_{r}}}+\frac{\textrm{e}^{\textrm{i}\frac{3\pi}{4}}\arctan\left(\frac{\textrm{e}^{\textrm{i}\frac{3\pi}{4}}k}{\sqrt{k^{2}_{\Delta}+\textrm{i}k^{2}_{r}}}\right)}{2\sqrt{k^{2}_{\Delta}+\textrm{i}k^{2}_{r}}}\right]_{0}^{k_{\textrm{F}}}=\frac{m\Delta}{\hbar^{2}k_{r}}\alpha_{1},\label{eq:D16}
\\
&\int_{0}^{k_{\textrm{F}}}\sin^{2}\eta_{2}\left(k\right)\textrm{d}k=\int_{0}^{k_{\textrm{F}}}\frac{\Delta^{2}}{\left(E_{r}-E_{k}\right)^{2}+\Delta^{2}}\textrm{d}k=\left(\frac{2m\Delta}{\hbar^{2}}\right)^{2}\int_{0}^{k_{\textrm{F}}}\frac{1}{\left(k^{2}_{r}-k^{2}\right)^{2}+k_{\Delta}^{4}}\textrm{d}k\nonumber\\&=-\frac{2m\Delta}{\hbar^{2}}\left[\frac{\textrm{e}^{\textrm{i}\frac{3\pi}{4}}\arctan\left(\frac{\textrm{e}^{\textrm{i}\frac{\pi}{4}}k}{\sqrt{k^{2}_{\Delta}-\textrm{i}k^{2}_{r}}}\right)}{2\sqrt{k^{2}_{\Delta}-\textrm{i}k^{2}_{r}}}+\frac{\textrm{e}^{\textrm{i}\frac{\pi}{4}}\arctan\left(\frac{\textrm{e}^{\textrm{i}\frac{3\pi}{4}}k}{\sqrt{k^{2}_{\Delta}+\textrm{i}k^{2}_{r}}}\right)}{2\sqrt{k^{2}_{\Delta}+\textrm{i}k^{2}_{r}}}\right]_{0}^{k_{\textrm{F}}}=\frac{m\Delta}{\hbar^{2}k_{r}}\alpha_{2}.\label{eq:D17}
\end{align}
For narrow virtual bound states close to the Fermi level, $\sqrt{2m\Delta/\hbar^{2}}=k_{\Delta}\ll k_{\textrm{F}}\approx k_{r}=\sqrt{2mE_{r}/\hbar^{2}}$, the factors $\alpha_{1}$ and $\alpha_{2}$ may be approximated as
\begin{align}
\alpha_{1}\approx&-\frac{1}{2}\ln{\left(\frac{\left(k_{r}-k_{\textrm{F}}\right)^{2}+\frac{k^{4}_{\Delta}}{4k^{2}_{r}}}{\left(k_{r}+k_{\textrm{F}}\right)^{2}+\frac{k^{4}_{\Delta}}{4k^{2}_{r}}}\right)},\label{eq:D18}
\\
\alpha_{2}\approx&\eta_{2}\left(k_{\textrm{F}}\right).\label{eq:D19}
\end{align}
When calculating the DMI in Ref.~\cite{levy1981anisotropy}, the integral over $k$ was rewritten to an integral over $\eta_{2}$ while keeping the $k$ dependence only in the phase shifts. This method works for the calculation of $\alpha_{2}$ (here we included $k_{r}$ in the prefactor instead of $k_{\textrm{F}}$ for increased numerical accuracy), but a different approximation is required for $\alpha_{1}$ which is sharply peaked at $k_{\textrm{F}}=k_{r}$.

In summary, the energy correction including scattering off the non-magnetic impurity may be written as
\begin{align}
E^{(2)}_{(1)}=&-\frac{45\pi\Gamma^{2}\Delta}{32E^{2}_{\textrm{F}}k^{2}_{\textrm{F}}k_{r}}\frac{\sin\left(k_{\textrm{F}}\left(R_{a}+R_{b}+R_{ab}\right)\right)\alpha_{1}+\left[\cos\left(k_{\textrm{F}}\left(R_{a}+R_{b}+R_{ab}\right)\right)-\cos\left(k_{\textrm{F}}\left(\left|R_{a}-R_{b}\right|+R_{ab}\right)\right)\right]\alpha_{2}}{R_{a}R_{b}R_{ab}}\nonumber\\&\times\boldsymbol{S}_{a}\cdot\boldsymbol{S}_{b}\left[3\left(\hat{\boldsymbol{R}_{a}}\cdot\hat{\boldsymbol{R}_{b}}\right)^{2}-1\right],\label{eq:D21}
\end{align}
where $E_{\textrm{F}}=\hbar^{2}k^{2}_{\textrm{F}}/(2m)$ and $N=k^{3}_{\textrm{F}}/(3\pi^2)$ was substituted, the latter assuming a single electron per site. The spatial dependence of this term is very similar to that of the DMI derived from the same model, although the prefactor does not contain the SOC since that was not included in the derivation here. However, the dependence on the spin directions is the same as in the RKKY interaction. This interaction term vanishes for certain angles between $\hat{\boldsymbol{R}_{a}}$ and $\hat{\boldsymbol{R}_{b}}$, but unlike for the DMI there is no global symmetry which could forbid the presence of this term when summing over the positions of all non-magnetic impurities.

To decrease the number of fitting parameters in the comparison with the first-principles results, we combined the $\sin\left(k_{\textrm{F}}\left(R_{a}+R_{b}+R_{ab}\right)\right)$ and $\cos\left(k_{\textrm{F}}\left(R_{a}+R_{b}+R_{ab}\right)\right)$ terms by introducing a phase factor $\phi'$, and neglected the $\cos\left(k_{\textrm{F}}\left(\left|R_{a}-R_{b}\right|+R_{ab}\right)\right)$ term, resulting in the expression used in the main text,
\begin{align}
E^{(2)}_{(1)}=&V'_0 \frac{\sin \left[ k_F \left( R_{A}+R_{B}+R_{AB} \right)+\phi' \right] }{\left(R_{A}\right)^{3}\left(R_{B}\right)^{3}R_{AB}} \left[3\left(\vec{R}_{A}\cdot\vec{R}_{B}\right)^{2}-\left(R_{A}\right)^{2}\left(R_{B}\right)^{2}\right]\boldsymbol{S}_{a}\cdot\boldsymbol{S}_{b}.
\end{align}
\twocolumngrid
\bibliography{references}

\end{document}